\documentclass[aps,eqsecnum,preprint,floats,epsf,epsfig,nofootinbib,letter]{revtex4}
\usepackage{epsfig}

\begin{document}
\def\be{\begin{eqnarray}}
\def\en{\end{eqnarray}}
\newcommand{\pslash}{\mbox{$\not{\hspace{-1.03mm}p}$}}        
\newcommand{\qslash}{\mbox{$\not{\hspace{-1.03mm}q}$}}        
\newcommand{\vslash}{\mbox{$\not{\hspace{-0.6mm}v}$}}        
\newcommand{\epsslash}{\mbox{$\not{\hspace{-1.03mm}\epsilon}$}} 
\newcommand{\Dslash}{\mbox{$\not{\hspace{-1.03mm}D}$}}        
\def\non{\nonumber}
\def\la{\langle}
\def\ra{\rangle}
\def\nc{N_c^{\rm eff}}
\def\vp{\varepsilon}
\def\drho{\bar\rho}
\def\deta{\bar\eta}
\def\CP{{\it CP}~}
\def\a{{\cal A}}
\def\B{{\cal B}}
\def\c{{\cal C}}
\def\d{{\cal D}}
\def\e{{\cal E}}
\def\p{{\cal P}}
\def\t{{\cal T}}
\def\S{{\cal S}}
\def\V{{\cal V}}
\def\R{{\cal R}}
\def\T{{\cal T}}
\def\U{{\cal U}}
\def\X{{\cal X}}
\def\up{\uparrow}
\def\dw{\downarrow}
\def\vma{{_{V-A}}}
\def\vpa{{_{V+A}}}
\def\smp{{_{S-P}}}
\def\spp{{_{S+P}}}
\def\J{{J/\psi}}
\def\ov{\overline}
\def\Lqcd{{\Lambda_{\rm QCD}}}
\def\pr{{Phys. Rev.}~}
\def\prl{{Phys. Rev. Lett.}~}
\def\pl{{Phys. Lett.}~}
\def\np{{Nucl. Phys.}~}
\def\zp{{Z. Phys.}~}
\def\lsim{ {\ \lower-1.2pt\vbox{\hbox{\rlap{$<$}\lower5pt\vbox{\hbox{$\sim$}
}}}\ } }
\def\gsim{ {\ \lower-1.2pt\vbox{\hbox{\rlap{$>$}\lower5pt\vbox{\hbox{$\sim$}
}}}\ } }

\font\el=cmbx10 scaled \magstep2{\obeylines\hfill December, 2006}

\vskip 1.5 cm

\centerline{\large\bf Strong Decays of Charmed Baryons}
\centerline{\large\bf in Heavy Hadron Chiral Perturbation Theory}
\bigskip
\centerline{\bf Hai-Yang Cheng$^{1}$ and Chun-Khiang Chua$^{2}$}
\medskip
\centerline{$^1$ Institute of Physics, Academia Sinica}
\centerline{Taipei, Taiwan 115, Republic of China}
\medskip
\centerline{$^2$ Department of Physics, Chung Yuan Christian
University} \centerline{Chung-Li, Taiwan 320, Republic of China}
\bigskip
\bigskip
\bigskip
\bigskip
\centerline{\bf Abstract}
\bigskip
\small

Strong decays of charmed baryons are analyzed in the framework of
heavy hadron chiral perturbation theory (HHChPT) in which heavy
quark symmetry and chiral symmetry are synthesized. HHChPT works
excellently for describing the strong decays of $s$-wave charmed
baryons. For $L=1$ orbitally excited states, two of the unknown
couplings, namely, $h_2$ and $h_{10}$, are determined from the
resonant $\Lambda_c^+\pi\pi$ mode produced in the
$\Lambda_c(2593)$ decay and the width of $\Sigma_c(2800)$,
respectively. Predictions for the strong decays of the $p$-wave
charmed baryon states $\Lambda_c(2625)$, $\Xi_c(2790)$ and
$\Xi_c(2815)$ are presented. Since the decay
$\Lambda_c(2593)^+\to\Lambda_c^+\pi\pi$  receives non-resonant
contributions, our value for $h_2$ is smaller than the previous
estimates. We also discuss the first positive-parity excited
charmed baryons. We conjecture that the charmed baryon
$\Lambda_c(2880)$ with $J^P=\frac52^+$ is an admixture of
$\Lambda_{c2}(\frac52^+)$ with and
$\tilde\Lambda''_{c3}(\frac52^+)$; both are $L=2$ orbitally
excited states. The potential model suggests $J^P=\frac52^-$ or
$\frac32^+$ for $\Lambda_c(2940)^+$. Measurements of the ratio of
$\Sigma_c^*\pi/\Sigma_c\pi$ will enable us to discriminate the
$J^P$ assignments for $\Lambda_c(2940)$. We advocate that the
$J^P$ quantum numbers of $\Xi_c(2980)$ and $\Xi_c(3077)$ are
$\frac12^+$ and $\frac52^+$, respectively. Under this $J^P$
assignment, it is easy to understand why $\Xi_c(2980)$ is broader
than $\Xi_c(3077)$.

\pagebreak

\section{Introduction}
In the past years many new excited charmed baryon states have been
discovered by BaBar, Belle and CLEO. In particular, $B$ factories
have provided a very rich source of charmed baryons both from $B$
decays and from the continuum $e^+e^-\to c\bar c$. A new era for
the charmed baryon spectroscopy is opened by the rich mass
spectrum and the relatively narrow widths of the excited states.
Experimentally and theoretically, it is important to identify the
quantum numbers of these new states and understand their
properties. Since the pseudoscalar mesons involved in the strong
decays of charmed baryons are soft, the charmed baryon system
offers an excellent ground for testing the ideas and predictions
of heavy quark symmetry of the heavy quarks and chiral symmetry of
the light quarks.

The observed mass spectra and decay widths of charmed baryons are
summarized in Tables \ref{tab:newbaryons} and \ref{tab:spectrum}.
Several new excited charmed baryon states  such as
$\Lambda_c(2765)^+,\Lambda_c(2880)^+,\Lambda_c(2940)^+$,
$\Xi_c(2815),\Xi_c(2980)$ and $\Xi_c(3077)$  have been measured
recently and they are still not on the particle listings of 2006
Review of Particle Physics by the Particle Data Group \cite{PDG}.
By now, the $J^P={1\over 2}^+$ and ${1\over 2}^-$ antitriplet
states: ($\Lambda_c^+$, $\Xi_c^+,\Xi_c^0)$, ($\Lambda_c(2593)^+$,
$\Xi_c(2790)^+,\Xi_c(2790)^0)$, and $J^P={1\over 2}^+$ and
${3\over 2}^+$ sextet states: ($\Omega_c,\Sigma_c,\Xi'_c$),
($\Omega_c^*,\Sigma_c^*,\Xi'^*_c$) are established. Notice that
except for the parity of the lightest $\Lambda_c^+$ and the
spin-parity of $\Lambda_c(2880)^+$, none of the other $J^P$
quantum numbers given in Table \ref{tab:spectrum} has been
measured. One has to rely on the quark model to determine the
$J^P$ assignments.

\begin{table}[b]
\caption{Masses (first entry) and decay widths (second entry) in
units of MeV for the excited  charmed baryons $\Lambda_c(2880)$,
$\Lambda_c(2940)$, $\Xi_c(2980)^{+,0}$, $\Xi_c(3077)^{+,0}$ and
$\Omega_c(2768)$.} \label{tab:newbaryons}
\begin{ruledtabular}
\begin{tabular}{c c c c c}
State & BaBar \cite{BaBar:Lamc2940,BaBar:Xic2980,BaBar:Omegacst}
& Belle \cite{Belle:Lamc2880,Belle:Xic2980} & CLEO \cite{CLEO:Lamc2880} & Average \\
\hline
 $\Lambda_c(2880)^+$ & $2881.9\pm0.1\pm0.5$ & $2881.2\pm0.2^{+0.4}_{-0.3}$ &
 $2882.5\pm2.2$ & $2881.5\pm0.3$  \\
 & $5.8\pm1.5\pm1.1$ & $5.5^{+0.7}_{-0.5}\pm0.4$ & $<8$ & $5.5\pm0.6$
 \\ \hline
 $\Lambda_c(2940)^+$ & $2939.8\pm1.3\pm1.0$ &
 $2937.9\pm1.0^{+1.8}_{-0.4}$ & & $2938.8\pm1.1$ \\
 & $17.5\pm5.2\pm5.9$ & $10\pm4\pm5$ & & $13.0\pm5.0$ \\ \hline
 $\Xi_c(2980)^+$ & $2967.1\pm1.9\pm1.0$ & $2978.5\pm2.1\pm2.0$  & &
 $2971.1\pm1.7$ \\
 & $23.6\pm2.8\pm1.3$ & $43.5\pm7.5\pm7.0$  & & $25.2\pm3.0$ \\
 \hline
 $\Xi_c(2980)^0$ & $$ & $2977.1\pm8.8\pm3.5$  & &
 $2977.1\pm9.5$ \\
 & $$ & $43.5~({\rm fixed})$  & & $43.5$ \\
 \hline
 $\Xi_c(3077)^+$ & $3076.4\pm0.7\pm0.3$ & $3076.7\pm0.9\pm0.5$ & &
 $3076.5\pm0.6$ \\
 & $6.2\pm1.6\pm0.5$ & $6.2\pm1.2\pm0.8$ & & $6.2\pm1.1$  \\
 \hline
 $\Xi_c(3077)^0$ & $$ & $3082.8\pm1.8\pm1.5$ & &
 $3082.8\pm2.3$ \\
 & $$ & $5.2\pm3.1\pm1.8$ & & $5.2\pm3.6$  \\
 \hline
 $\Omega_c(2768)^0$ & $2768.3\pm3.0$ & & & $2768.3\pm3.0$ \\
\end{tabular}
\end{ruledtabular}
\end{table}

\begin{table}[!]
\caption{Mass spectra and decay widths (in units of MeV) of
charmed baryons. Experimental values are taken from the Particle
Data Group \cite{PDG} and Table \ref{tab:newbaryons}.}
\label{tab:spectrum}
\begin{center}
\begin{tabular}{|c|c|ccc|c|c|c|c|} \hline \hline
~~State~~ & ~~$J^P$~~ &~$S_\ell$~ & ~$L_\ell$~ &
~$J_\ell^{P_\ell}$~ &
~~~~~~~~~Mass~~~~~~~~~ & ~~~~Width~~~~ &~Decay modes~\\
\hline
 $\Lambda_c^+$ & ${1\over 2}^+$ & 0 & 0 & $0^+$ & $2286.46\pm0.14$ & & weak  \\
 \hline
 $\Lambda_c(2593)^+$ & ${1\over 2}^-$ & 0 & 1 & $1^-$ & $2595.4\pm0.6$ &
 $3.6^{+2.0}_{-1.3}$ & $\Sigma_c\pi,\Lambda_c\pi\pi$ \\
 \hline
 $\Lambda_c(2625)^+$ & ${3\over 2}^-$ & 0 & 1 & $1^-$ & $2628.1\pm0.6$ &
 $<1.9$ & $\Lambda_c\pi\pi,\Sigma_c\pi$ \\
 \hline
 $\Lambda_c(2765)^+$ & $?^?$ & ? & ? & $?$ & $2766.6\pm2.4$ & $50$ & $\Sigma_c\pi,\Lambda_c\pi\pi$ \\
 \hline
 $\Lambda_c(2880)^+$ & ${5\over 2}^+$ & ? & ? & ? & $2881.5\pm0.3$ & $5.5\pm0.6$
 & $\Sigma_c^{(*)}\pi,\Lambda_c\pi\pi,D^0p$ \\
 \hline
 $\Lambda_c(2940)^+$ & $?^?$ & ? & ? & $?$ & $2938.8\pm1.1$ & $13.0\pm5.0$ &
 $\Sigma_c^{(*)}\pi,\Lambda_c\pi\pi,D^0p$ \\ \hline
 $\Sigma_c(2455)^{++}$ & ${1\over 2}^+$ & 1 & 0 & $1^+$ & $2454.02\pm0.18$ &
 $2.23\pm0.30$ & $\Lambda_c\pi$ \\
 \hline
 $\Sigma_c(2455)^{+}$ & ${1\over 2}^+$ & 1 & 0 & $1^+$ & $2452.9\pm0.4$ &
 $<4.6$ & $\Lambda_c\pi$\\
 \hline
 $\Sigma_c(2455)^{0}$ & ${1\over 2}^+$ & 1 & 0 & $1^+$ & $2453.76\pm0.18$
 & $2.2\pm0.4$ & $\Lambda_c\pi$ \\
 \hline
 $\Sigma_c(2520)^{++}$ & ${3\over 2}^+$ & 1 & 0 & $1^+$ & $2518.4\pm0.6$
 & $14.9\pm1.9$ & $\Lambda_c\pi$\\
 \hline
 $\Sigma_c(2520)^{+}$ & ${3\over 2}^+$ & 1 & 0 & $1^+$ & $2517.5\pm2.3$
 & $<17$ & $\Lambda_c\pi$ \\
 \hline
 $\Sigma_c(2520)^{0}$ & ${3\over 2}^+$ & 1 & 0 & $1^+$ & $2518.0\pm0.5$
 & $16.1\pm2.1$ & $\Lambda_c\pi$ \\
 \hline
 $\Sigma_c(2800)^{++}$ & ${3\over 2}^-$? & 1 & 1 & $2^-$ & $2801^{+4}_{-6}$ & $75^{+22}_{-17}$ &
 $\Lambda_c\pi,\Sigma_c^{(*)}\pi,\Lambda_c\pi\pi$ \\
 \hline
 $\Sigma_c(2800)^{+}$ & ${3\over 2}^-$? & 1 & 1 & $2^-$ & $2792^{+14}_{-5}$ & $62^{+60}_{-40}$ &
 $\Lambda_c\pi,\Sigma_c^{(*)}\pi,\Lambda_c\pi\pi$ \\
 \hline
 $\Sigma_c(2800)^{0}$ & ${3\over 2}^-$? &1 & 1 & $2^-$ & $2802^{+4}_{-7}$ & $61^{+28}_{-18}$ &
 $\Lambda_c\pi,\Sigma_c^{(*)}\pi,\Lambda_c\pi\pi$\\
 \hline
 $\Xi_c^+$ & ${1\over 2}^+$ & 0 & 0 & $0^+$ & $2467.9\pm0.4$ & & weak \\ \hline
 $\Xi_c^0$ & ${1\over 2}^+$ & 0 & 0 & $0^+$ & $2471.0\pm0.4$ & & weak \\ \hline
 $\Xi'^+_c$ & ${1\over 2}^+$ & 1 & 0 & $1^+$ & $2575.7\pm3.1$ & & $\Xi_c\gamma$ \\ \hline
 $\Xi'^0_c$ & ${1\over 2}^+$ & 1 & 0 & $1^+$ & $2578.0\pm2.9$ & & $\Xi_c\gamma$ \\ \hline
 $\Xi_c(2645)^+$ & ${3\over 2}^+$ & 1 & 0 & $1^+$ & $2646.6\pm1.4$ & $<3.1$ & $\Xi_c\pi$ \\
 \hline
 $\Xi_c(2645)^0$ & ${3\over 2}^+$ & 1 & 0 & $1^+$ & $2646.1\pm1.2$ & $<5.5$ & $\Xi_c\pi$ \\
 \hline
 $\Xi_c(2790)^+$ & ${1\over 2}^-$ & 0 & 1 & $1^-$ & $2789.2\pm3.2$ & $<15$ & $\Xi'_c\pi$\\
 \hline
 $\Xi_c(2790)^0$ & ${1\over 2}^-$ & 0 & 1 & $1^-$ & $2791.9\pm3.3$ & $<12$ & $\Xi'_c\pi$ \\
 \hline
 $\Xi_c(2815)^+$ & ${3\over 2}^-$ & 0 & 1 & $1^-$ & $2816.5\pm1.2$ & $<3.5$ & $\Xi^*_c\pi,\Xi_c\pi\pi,\Xi_c'\pi$ \\
 \hline
 $\Xi_c(2815)^0$ & ${3\over 2}^-$ & 0 & 1 & $1^-$ & $2818.2\pm2.1$ & $<6.5$ & $\Xi^*_c\pi,\Xi_c\pi\pi,\Xi_c'\pi$ \\
 \hline
 $\Xi_c(2980)^+$ & $?^?$ & ? & ? & $?$ & $2971.1\pm1.7$ & $25.2\pm3.0$
 & see Table \ref{tab:Xic} \\
 \hline
 $\Xi_c(2980)^0$ & $?^?$ & ? & ? & $?$ & $2977.1\pm9.5$ & $43.5$
 & see Table \ref{tab:Xic} \\
 \hline
 $\Xi_c(3077)^+$ & $?^?$ & ? & ? & $?$ & $3076.5\pm0.6$ & $6.2\pm1.1$ &
 see Table \ref{tab:Xic}  \\
 \hline
 $\Xi_c(3077)^0$ & $?^?$ & ? & ? & $?$ & $3082.8\pm2.3$ & $5.2\pm3.6$
 & see Table \ref{tab:Xic} \\
 \hline
 $\Omega_c^0$ & ${1\over 2}^+$ & 1 & 0 & $1^+$ & $2697.5\pm2.6$ & & weak \\
 \hline
 $\Omega_c(2768)^0$ & ${3\over 2}^+$ & 1 & 0 & $1^+$ & $2768.3\pm3.0$ & & $\Omega_c\gamma$ \\
 \hline \hline
\end{tabular}
\end{center}
\end{table}

This work is organized as follows. In Sec. II, the experimental
status of the charmed baryon spectroscopy is reviewed. The
$p$-wave charmed baryons and the first positive parity excitations
are discussed. In Sec. III we first present the relevant chiral
Lagrangians which combine heavy quark and chiral symmetries. Then
we proceed to the phenomenological implications to the strong
decays of $s$-wave and $p$-wave charmed baryons as well as first
positive parity excited charmed baryon states. Conclusions are
presented in Sec. IV.

\section{Spectroscopy}

Charmed baryon spectroscopy provides an ideal place for studying
the dynamics of the light quarks in the environment of a heavy
quark. The charmed baryon of interest contains a charmed quark and
two light quarks, which we will often refer to as a diquark. Each
light quark is a triplet of the flavor SU(3). Since ${\bf 3}\times
{\bf 3}={\bf \bar 3}+{\bf 6}$, there are two different SU(3)
multiplets of charmed baryons: a symmetric sextet {\bf 6} and an
antisymmetric antitriplet ${\bf \bar 3}$. The spin-flavor-space
wave functions of baryons are totally symmetric since the color
wave function is totally antisymmetric. For the ground-state
$s$-wave baryons in the quark model, the symmetries in the flavor
and spin of the diquarks are thus correlated. Consequently, the
diquark in the flavor-symmetric sextet has spin 1, while the
diquark in the flavor-antisymmetric antitriplet has spin 0. When
the diquark combines with the charmed quark, the sextet contains
both spin-$\frac12$ and spin-$\frac32$ charmed baryons. However,
the antitriplet contains only spin-$\frac12$ ones. More
specifically, the $\Lambda_c^+$, $\Xi_c^+$ and $\Xi_c^0$ form a
${\bf \bar 3}$ representation and they all decay weakly. The
$\Omega_c^0$, $\Xi'^+_c$, $\Xi'^0_c$ and $\Sigma_c^{++,+,0}$ form
a {\bf 6} representation; among them, only $\Omega_c^0$ decays
weakly. Note that we follow the Particle Data Group (PDG)
\cite{PDG} to use a prime to distinguish the $\Xi_c$ in the {\bf
6} from the one in the ${\bf \bar 3}$.

The lowest-lying orbitally excited baryon states are the $p$-wave
charmed baryons with their quantum numbers listed in Table
\ref{tab:pwave}. Although the separate spin angular momentum
$S_\ell$ and orbital angular momentum $L_\ell$ of the light
degrees of freedom are not well defined, they are included for
guidance from the quark model. In the heavy quark limit, the spin
of the charmed quark $S_c$ and the total angular momentum of the
two light quarks $J_\ell=S_\ell+L_\ell$ are separately conserved.
It is convenient to use them to enumerate the spectrum of states.
There are two types of $L_\ell=1$ orbital excited charmed baryon
states: states with  the unit of orbital angular momentum between
the diquark and the charmed quark, and states with the unit of
orbital angular momentum between the two light quarks. The orbital
wave function of the former (latter) is symmetric (antisymmetric)
under the exchange of two light quarks. To see this, one can
define two independent relative momenta ${\bf k}={1\over 2}({\bf
p}_1-{\bf p}_2)$ and ${\bf K}={1\over 2}({\bf p}_1+{\bf p}_2-2{\bf
p}_c)$ from the two light quark momenta ${\bf p}_1$, ${\bf p}_2$
and the heavy quark momentum ${\bf p}_c$. (In the heavy quark
limit, ${\bf p}_c$ can be set to zero.) Denoting the quantum
numbers $L_k$ and $L_K$ as the eigenvalues of ${\bf L}_k^2$ and
${\bf L}_K^2$,\footnote{In the notation of \cite{Capstick}, $L_k$
and $L_K$ correspond to $\ell_\rho$ and $\ell_\lambda$,
respectively.}
the $k$-orbital momentum $L_k$ describes relative orbital
excitations of the two light quarks, and the $K$-orbital momentum
$L_K$ describes orbital excitations of the center of the mass of
the two light quarks relative to the heavy quark \cite{Korner94}.
The $p$-wave heavy baryon can be either in the $(L_k=0,L_K=1)$
$K$-state or the $(L_k=1,L_K=0)$ $k$-state. It is obvious that the
orbital $K$-state ($k$-state) is symmetric (antisymmetric) under
the interchange of ${\bf p}_1$ and ${\bf p}_2$.

\begin{table}[t]
\caption{The $p$-wave charmed baryons and their quantum numbers,
where $S_\ell$ ($J_\ell$) is the total spin (angular momentum) of
the two light quarks. The quantum number in the subscript labels
$J_\ell$, while the quantum number in parentheses is referred to
the spin of the baryon. In the quark model, the upper (lower)
eight multiplets have even (odd) orbital wave functions under the
permutation of the two light quarks. That is, $L_\ell$ for the
former is referred to the orbital angular momentum between the
diquark and the charmed quark, while $L_\ell$ for the latter is
the orbital angular momentum between the two light quarks. The
explicit quark model wave functions for $p$-wave charmed baryons
can be found in \cite{Pirjol}. The states antisymmetric in orbital
wave functions are denoted by a tilde, while the superscript prime
is reserved for the $\Xi_c$ charmed baryons to distinguish between
the sextet and antitriplet SU(3) flavor states.   }
\label{tab:pwave}
\begin{center}
\begin{tabular}{|c|cccc||c|cccc|} \hline
~~~~~State~~~~~ & SU(3)$_F$ & ~~$S_\ell$~~ & ~~$L_\ell$~~&
~~$J_\ell^{P_\ell}$~~ & ~~~~~State~~~~~ & SU(3)$_F$ & ~~$S_\ell$~~
& ~~$L_\ell$~~&
~~$J_\ell^{P_\ell}$~ \\
 \hline
 $\Lambda_{c1}({1\over 2}^-,{3\over 2}^-)$ & ${\bf \bar 3}$ & 0 & 1 &
 $1^-$ & $\Xi_{c1}({1\over 2}^-,{3\over 2}^-)$ & ${\bf \bar 3}$ & 0 & 1 & $1^-$ \\
 $\Sigma_{c0}({1\over 2}^-)$ & ${\bf 6}$ & 1 & 1& $0^-$ &
 $\Xi'_{c0}({1\over 2}^-)$ & ${\bf 6}$ & 1 & 1& $0^-$ \\
 $\Sigma_{c1}({1\over 2}^-,{3\over 2}^-)$ & ${\bf 6}$ & 1 & 1 & $1^-$
 &  $\Xi'_{c1}({1\over 2}^-,{3\over 2}^-)$ & ${\bf 6}$ & 1 & 1 &
 $1^-$\\
 $\Sigma_{c2}({3\over 2}^-,{5\over 2}^-)$ & ${\bf 6}$ & 1 & 1 & $2^-$
 &  $\Xi'_{c2}({3\over 2}^-,{5\over 2}^-)$ & ${\bf 6}$ & 1 & 1 &
 $2^-$\\  \hline
 $\tilde \Sigma_{c1}({1\over 2}^-,{3\over 2}^-)$ & ${\bf 6}$ & 0 & 1 & $1^-$
 &  $\tilde\Xi'_{c1}({1\over 2}^-,{3\over 2}^-)$ & ${\bf 6}$ & 0 & 1 & $1^-$
 \\
 $\tilde\Lambda_{c0}({1\over 2}^-)$ & ${\bf \bar 3}$ & 1 & 1 & $0^-$ &
 $\tilde\Xi_{c0}({1\over 2}^-)$ & ${\bf \bar 3}$ & 1 & 1 & $0^-$  \\
 $\tilde\Lambda_{c1}({1\over 2}^-,{3\over 2}^-)$ & ${\bf \bar 3}$ & 1 & 1 & $1^-$
 &  $\tilde\Xi_{c1}({1\over 2}^-,{3\over 2}^-)$ & ${\bf \bar 3}$ & 1 & 1 &
 $1^-$ \\
 $\tilde\Lambda_{c2}({3\over 2}^-,{5\over 2}^-)$ & ${\bf \bar 3}$ & 1 & 1 & $2^-$
 &  $\tilde\Xi_{c2}({3\over 2}^-,{5\over 2}^-)$ & ${\bf \bar 3}$ & 1 & 1 & $2^-$ \\
 \hline
\end{tabular}
\end{center}
\end{table}

There are seven lowest-lying $p$-wave $\Lambda_c$ arising from
combining the charmed quark spin $S_c$ with light constituents in
$J_\ell^{P_\ell}=1^-$ state: three $J^P=\frac12^-$ states, three
$J^P=\frac32^-$ states and one $J^P=\frac52^-$ state. They form
three doublets $\Lambda_{c1}({1\over 2},{3\over
2}),\tilde\Lambda_{c1}({1\over 2},{3\over
2}),\tilde\Lambda_{c2}({3\over 2},{5\over 2})$ and one singlet
$\tilde\Lambda_{c0}({1\over 2})$, where we have used a tilde to
denote the multiplets antisymmetric in the orbital wave functions
under the exchange of two light quarks. In terms of $K$- and
$k$-states introduced above, the doublets
$\Lambda_{c1},\tilde\Lambda_{c1},\tilde\Lambda_{c2}$ and the
singlet $\tilde\Lambda_{c0}$ are sometimes denoted by
$\Lambda_{cK1},\Lambda_{ck1},\Lambda_{ck2}$ and $\Lambda_{ck0}$,
respectively, in the literature \cite{Korner94}. Quark models
\cite{Capstick} indicate that the untilde states for $\Lambda$-
and $\Sigma$-type charmed baryons with symmetric orbital wave
functions lie about 150 MeV below the tilde ones. The two states
in each doublet with $J=J_\ell\pm{1\over 2}$ are nearly
degenerate; their masses split only by a chromomagnetic
interaction.

Since the spin-parity of the newly measured $\Lambda_c(2880)^+$
was recently pinned down to be $\frac52^+$ by Belle
\cite{Belle:Lamc2880}, we shall briefly discuss the
positive-parity excitations of charmed baryons. Referring to the
orbital angular momentum quantum numbers $L_K$ and $L_k$, the
first positive parity excitations are those states with
$L_K+L_k=2$. For $L_K=2,L_k=0,L=2$ or $L_K=0,L_k=2,L=2$, there is
one multiplet for positive-parity excited $\Lambda_c$ and three
multiplets for $\Sigma_c$ as tabulated in Table
\ref{tab:pp}.\footnote{Strictly spaking, there are two multiplets
for positive-parity excited $\Lambda_c$ and six multiplets for
$\Sigma_c$ coming from two different orbital states $L_K=2,L_k=0$
and $L_K=0,L_k=2$. For simplicity, here we will not distinguish
between them.}
The orbital states of these multiplets are symmetric under the
interchange of the two light quarks. For the case of $L_K=L_k=1$,
the total orbital angular momentum $L_\ell$ of the diquark is 2, 1
or 0. Since the orbital states are antisymmetric under the
interchange of two light quarks, we shall use a tilde to denote
the $L_K=L_k=1$ states.  The Fermi-Dirac statistics for baryons
yields seven more multiplets for positive-parity excited
$\Lambda_c$ states and three more multiplets for $\Sigma_c$
baryons. The reader is referred to \cite{Hussain} for more
details.

\begin{table}[t]
\caption{The first positive-parity excitations of charmed baryons
and their quantum numbers. States with antisymmetric orbital wave
functions (i.e. $L_K=L_k=1$) under the interchange of two light
quarks are denoted by a tilde. A prime is used to distinguish
between the sextet and antitriplet SU(3) flavor states of the
excited $\Xi_c$. } \label{tab:pp}
\begin{center}
\begin{tabular}{|c|cccc||c|cccc|} \hline
~~~~~State~~~~~ & SU(3)$_F$ & ~~$S_\ell$~~ & ~~$L_\ell$~~&
~~$J_\ell^{P_\ell}$~~ & ~~~~~State~~~~~ & SU(3)$_F$ & ~~$S_\ell$~~
& ~~$L_\ell$~~&
~~$J_\ell^{P_\ell}$~ \\
 \hline
 $\Lambda_{c2}({3\over 2}^+,{5\over 2}^+)$ & ${\bf \bar 3}$ & 0 & 2 &
 $2^+$ & $\Sigma_{c1}({1\over 2}^+,{3\over 2}^+)$ & ${\bf 6}$ & 1 & 2 & $1^+$ \\
 $\tilde\Lambda_{c1}({1\over 2}^+,\frac32^+)$ & ${\bf \bar 3}$ & 1 & 0 & $1^+$ &
 $\Sigma_{c2}({3\over 2}^+,{5\over 2}^+)$ & ${\bf 6}$ & 1 & 2 & $2^+$ \\
 $\tilde\Lambda'_{c0}({1\over 2}^+)$ & ${\bf \bar 3}$ & 1 & 1 & $0^+$ &
 $\Sigma_{c3}({5\over 2}^+,{7\over 2}^+)$ & ${\bf 6}$ & 1 & 2 & $3^+$ \\
 $\tilde\Lambda'_{c1}({1\over 2}^+,{3\over 2}^+)$ & ${\bf \bar 3}$ & 1 & 1 & $1^+$ &
 $\tilde\Sigma_{c0}({1\over 2}^+)$ & ${\bf 6}$ & 0 & 0 & $0^+$ \\
 $\tilde\Lambda'_{c2}({3\over 2}^+,{5\over 2}^+)$ & ${\bf \bar 3}$ & 1 & 1 & $2^+$ &
 $\tilde\Sigma_{c1}({1\over 2}^+,{3\over 2}^+)$ & ${\bf 6}$ & 0 & 1 & $1^+$ \\
 $\tilde\Lambda''_{c1}({1\over 2}^+,{3\over 2}^+)$ & ${\bf \bar 3}$ & 1 & 2 & $1^+$ &
 $\tilde\Sigma_{c2}({3\over 2}^+,{5\over 2}^+)$ & ${\bf 6}$ & 0 & 2 & $2^+$ \\
 $\tilde\Lambda''_{c2}({3\over 2}^+,{5\over 2}^+)$ & ${\bf \bar 3}$ & 1 & 2 & $2^+$ &
 & & & & \\
 $\tilde\Lambda''_{c3}({5\over 2}^+,{7\over 2}^+)$ & ${\bf \bar 3}$ & 1 & 2 & $3^+$ &
 & & & & \\
 \hline
 $\Xi_{c2}({3\over 2}^+,{5\over 2}^+)$ & ${\bf \bar 3}$ & 0 & 2 &
 $2^+$ & $\Xi'_{c1}({1\over 2}^+,{3\over 2}^+)$ & ${\bf 6}$ & 1 & 2 & $1^+$ \\
 $\tilde\Xi_{c1}({1\over 2}^+,\frac32^+)$ & ${\bf \bar 3}$ & 1 & 0 & $1^+$ &
 $\Xi'_{c2}({3\over 2}^+,{5\over 2}^+)$ & ${\bf 6}$ & 1 & 2 & $2^+$ \\
 $\tilde\Xi''_{c0}({1\over 2}^+)$ & ${\bf \bar 3}$ & 1 & 1 & $0^+$ &
 $\Xi'_{c3}({5\over 2}^+,{7\over 2}^+)$ & ${\bf 6}$ & 1 & 2 & $3^+$ \\
 $\tilde\Xi''_{c1}({1\over 2}^+,{3\over 2}^+)$ & ${\bf \bar 3}$ & 1 & 1 & $1^+$ &
 $\tilde\Xi'_{c0}({1\over 2}^+)$ & ${\bf 6}$ & 0 & 0 & $0^+$ \\
 $\tilde\Xi''_{c2}({3\over 2}^+,{5\over 2}^+)$ & ${\bf \bar 3}$ & 1 & 1 & $2^+$ &
 $\tilde\Xi'_{c1}({1\over 2}^+,{3\over 2}^+)$ & ${\bf 6}$ & 0 & 1 & $1^+$ \\
 $\tilde\Xi'''_{c1}({1\over 2}^+,{3\over 2}^+)$ & ${\bf \bar 3}$ & 1 & 2 & $1^+$ &
 $\tilde\Xi'_{c2}({3\over 2}^+,{5\over 2}^+)$ & ${\bf 6}$ & 0 & 2 & $2^+$ \\
 $\tilde\Xi'''_{c2}({3\over 2}^+,{5\over 2}^+)$ & ${\bf \bar 3}$ & 1 & 2 & $2^+$ &
 & & & & \\
 $\tilde\Xi'''_{c3}({5\over 2}^+,{7\over 2}^+)$ & ${\bf \bar 3}$ & 1 & 2 & $3^+$ &
 & & & & \\
 \hline
\end{tabular}
\end{center}
\end{table}

In the following we discuss some of the new excited charmed baryon
states:
\subsection{$\Lambda_c$}

$\Lambda_c(2593)^+$ and $\Lambda_c(2625)^+$ form a doublet
$\Lambda_{c1}({1\over 2}^-,{3\over 2}^-)$ \cite{Cho}. The dominant
decay mode is $\Sigma_c\pi$ in an $S$ wave for
$\Lambda_{c1}({1\over 2}^-)$ and $\Lambda_c\pi\pi$ in a $P$ wave
for $\Lambda_{c1}({3\over 2}^-)$. (The two-body mode $\Sigma_c\pi$
is a $D$-wave in $\Lambda_c(\frac32^-)$ decay.) This explains why
the width of $\Lambda_c(2625)^+$ is narrower than that of
$\Lambda_c(2593)^+$.

$\Lambda_c(2765)^+$ is a broad state ($\Gamma\approx 50$ MeV)
first seen in $\Lambda_c^+\pi^+\pi^-$ by CLEO
\cite{CLEO:Lamc2880}. It appears to resonate through $\Sigma_c$
and probably also $\Sigma_c^*$. However, whether it is a
$\Lambda_c^+$ or a $\Sigma_c^+$ or whether the width might be due
to overlapping states are not known. The Skyrme model \cite{Oh}
and the quark model \cite{Capstick} suggest a $J^P=\frac12^+$
$\Lambda_c$ state with a mass 2742 and 2775 MeV, respectively.
Therefore, $\Lambda_c(2765)^+$ could be a first positive-parity
excitation of $\Lambda_c$.

The state $\Lambda_c(2880)^+$ first observed by CLEO
\cite{CLEO:Lamc2880} in $\Lambda_c^+\pi^+\pi^-$ was also seen by
BaBar in the $D^0p$ spectrum \cite{BaBar:Lamc2940}. It was
originally conjectured that, based on its narrow width,
$\Lambda_c(2880)^+$ might be a $\tilde\Lambda^+_{c0}({1\over
2}^-)$ state \cite{CLEO:Lamc2880}. Recently, Belle has studied the
experimental constraint on the $J^P$ quantum numbers of
$\Lambda_c(2880)^+$ \cite{Belle:Lamc2880}. The angular analysis of
$\Lambda_c(2880)^+\to\Sigma_c^{0,++}\pi^\pm$ indicates that
$J=\frac52$ is favored over $J=\frac12$ or $\frac32$, while the
study of the resonant structure of
$\Lambda_c(2880)^+\to\Lambda_c^+\pi^+\pi^-$ implies the existence
of the $\Sigma_c^*\pi$ intermediate states and
$\Gamma(\Sigma_c^*\pi^\pm)/\Gamma(\Sigma_c\pi^\pm)=(24.1\pm6.4^{+1.1}_{-4.5})\%$.
This value is in agreement with heavy quark symmetry predictions
\cite{IW} and favors the $\frac52^+$ over the $\frac52^-$
assignment. We shall return back to this point in Sec. III.C. It
is interesting to notice that, based on the diquark idea, the
quantum numbers $J^P=\frac52^+$ have already been predicted in
\cite{Selem} for the $\Lambda_c(2880)$ before the Belle
experiment.

The highest $\Lambda_c(2940)^+$ was first discovered by BaBar in
the $D^0p$ decay mode \cite{BaBar:Lamc2940} and  confirmed by
Belle in the decays $\Sigma_c^0\pi^+,\Sigma_c^{++}\pi^-$ which
subsequently decay into $\Lambda_c^+\pi^+\pi^-$
\cite{Belle:Lamc2880,Mizuk}. Since the mass of $\Lambda_c(2940)^+$
is barely below the threshold of $D^{*0}p$, this observation has
motivated the authors of \cite{He} to suggest an exotic molecular
state of $D^{*0}$ and $p$ with a binding energy of order 6 MeV for
$\Lambda_c(2940)^+$. Its quantum numbers $J^P$ could be
$\frac32^+$ or $\frac52^-$ as suggested by the quark model
calculation \cite{Capstick}.

\subsection{$\Sigma_c$}
The highest isotriplet charmed baryons $\Sigma_c(2800)^{++,+,0}$
decaying to $\Lambda_c^+\pi$ were first measured by Belle
\cite{Belle:Sigc2800}.  They are most likely to be the
$J^P=\frac32^-$ $\Sigma_{c2}$ states because the
$\Sigma_{c2}({3\over 2}^-)$ baryon decays principally into the
$\Lambda_c\pi$ system in a $D$-wave, while $\Sigma_{c1}({3\over
2}^-)$ decays mainly to the two pion system $\Lambda_c\pi\pi$ in a
$P$-wave. The state $\Sigma_{c0}({1\over 2}^-)$ can decay into
$\Lambda_c\pi$ in an $S$-wave, but it is very broad with width of
order 406 MeV (see Sec. III.C).

\subsection{$\Xi_c$}

The states $\Xi_c(2790)$ and $\Xi_c(2815)$ form a doublet
$\Xi_{c1}({1\over 2}^-,{3\over 2}^-)$. Since the diquark
transition $1^-\to 0^++\pi$ is prohibited, $\Xi_{c1}({1\over
2}^-,{3\over 2}^-)$ cannot decay to $\Xi_c\pi$. The dominant decay
mode is $[\Xi'_c\pi]_S$ for $\Xi_{c1}({1\over 2}^-)$ and
$[\Xi_c^*\pi]_S$ for $\Xi_{c1}({3\over 2}^-)$ where $\Xi_c^*$
stands for $\Xi_c(2645)$.

The new charmed strange baryons $\Xi_c(2980)^+$ and
$\Xi_c(3077)^+$ that decay into $\Lambda_c^+K^-\pi^+$ were first
observed by Belle \cite{Belle:Xic2980} and confirmed by BaBar
\cite{BaBar:Xic2980}. In the recent BaBar measurement
\cite{BaBar:Xic2980}, the $\Xi_{c}(2980)^+$ is found to decay
resonantly through the intermediate state $\Sigma_c(2455)^{++}K^-$
with $4.9\,\sigma$ significance and non-resonantly to
$\Lambda_c^+K^-\pi^+$ with $4.1\,\sigma$ significance. With
$5.8\,\sigma$ significance, the $\Xi_c(3077)^+$ is found to decay
resonantly through $\Sigma_c(2455)^{++}K^-$, and with
$4.6\,\sigma$ significance, it is found to decay through
$\Sigma_c(2520)^{++}K^-$. The significance of the signal for the
non-resonant decay $\Xi_c(3077)^+\rightarrow\Lambda_c^+K^-\pi^+$
is $1.4\,\sigma$.

\subsection{$\Omega_c$}
At last, the $J^P=\frac32^+$ $\Omega_c(2768)$ charmed baryon was
recently observed by BaBar in the decay
$\Omega_c(2768)^0\to\Omega_c^0\gamma$ \cite{BaBar:Omegacst}. With
this new observation, the $\frac32^+$ sextet is finally completed.
However, it will be very difficult to measure the electromagnetic
decay rate  because the width of $\Omega_c^*$, which is predicted
to be of order 0.9 keV \cite{ChengSU3}, is too narrow to be
experimentally resolvable.

\section{Strong decays}

Due to the rich mass spectrum and the relatively narrow widths of
the excited states, the charmed baryon system offers an excellent
ground for testing the ideas and predictions of heavy quark
symmetry and light flavor SU(3) symmetry. The pseudoscalar mesons
involved in the strong decays of charmed baryons such as
$\Sigma_c\to\Lambda_c\pi$ are soft. Therefore, heavy quark
symmetry of the heavy quark and chiral symmetry of the light
quarks will have interesting implications for the low-energy
dynamics of heavy baryons interacting with the Goldstone bosons.

The strong decays of charmed baryons are most conveniently
described by the heavy hadron chiral Lagrangians in which heavy
quark symmetry and chiral symmetry are incorporated
\cite{Yan,Wise}. The Lagrangian involves two coupling constants
$g_1$ and $g_2$ for $P$-wave transitions between $s$-wave and
$s$-wave baryons \cite{Yan}, six couplings $h_{2}-h_7$ for the
$S$-wave transitions between $s$-wave and $p$-wave baryons, and
eight couplings $h_{8}-h_{15}$ for the $D$-wave transitions
between $s$-wave and $p$-wave baryons \cite{Pirjol}.

Since the general chiral Lagrangian for heavy baryons coupling to
the pseudoscalar mesons can be expressed compactly in terms of
superfields, we first introduce the superfields for $s$-wave
baryons given by
 \be  \label{eq:S}
 \S_\mu^{ij} &=& \frac{1+\vslash}{2} B^{*ij}_{6\mu}+
\frac{1}{\sqrt3}(\gamma_\mu+v_\mu)\gamma_5\frac{1+\vslash}{2}
B_6^{ij}\,, \non \\
 \bar \S_\mu^{ij} &=& \bar B^{*ij}_{6\mu}\frac{1+\vslash}{2}
 -\frac{1}{\sqrt3}\bar B_6^{ij}\frac{1+\vslash}{2}
\gamma_5(\gamma_\mu+v_\mu) \,, \non \\
\T_i &=& \frac{1+\vslash}{2}\left(\Xi_c^0, -\Xi_c^+, \Lambda_c^+
\right)_i  = \frac12 \epsilon_{ijk}\frac{1+\vslash}{2}(B_{\bar
3})_{jk}\,.
 \en
where the matrices $B_6$ and $B^*_{6\mu}$ are defined in
\cite{Yan}
 \be  \label{3} (B_6)_{ij} = \left(
\begin{array}{ccc}
\Sigma_c^{++} & \frac{1}{\sqrt2}\Sigma_c^+ & \frac{1}{\sqrt2}\Xi'^+_c \\
\frac{1}{\sqrt2}\Sigma_c^+ & \Sigma_c^0 & \frac{1}{\sqrt2}\Xi'^0_c \\
\frac{1}{\sqrt2}\Xi_c^{+'} & \frac{1}{\sqrt2}\Xi'^0_c & \Omega_c^0
\end{array} \right)_{ij}, \qquad
 (B_{\bar 3})_{ij} = \left( \begin{array}{ccc}
0 & \Lambda_c^+ & \Xi_c^+ \\
-\Lambda_c^+ & 0 & \Xi_c^0 \\
-\Xi_c^+ & -\Xi_c^0 & 0
\end{array} \right)_{ij}.
 \en
The superfield for $p$-wave ${\bf \bar 3}$ multiplets symmetric in
orbital wave functions such as $\Lambda_{c1}({1\over 2}^-,{3\over
2}^-)$ and $\Xi_{c1}({1\over 2}^-,{3\over 2}^-)$ is given by
 \be \label{eq:R} \R_\mu^i =
\frac{1}{\sqrt3}(\gamma_\mu+v_\mu)\gamma_5 R^i + R^{*i}_\mu,
 \en
with
 \be   \label{}
 R_i = \frac{1+\vslash}{2}\left(\Xi_{c_1}^0,
-\Xi_{c_1}^+, \Lambda_{c_1}^+\right)_i\,,\qquad R_\mu^{*i} =
\frac{1+\vslash}{2}\left(\Xi_{c_1}^{*0},
 -\Xi_{c_1}^{*+}, \Lambda_{c_1}^{*+}\right)_{\mu i}\,,
 \en
where $\B_c^*$ denotes a spin-${3\over 2}$ charmed baryon. Note
that $v\cdot \S=v\cdot\R=0$.

There are three other $p$-wave sextet multiplets with
antisymmetric orbital states and with quantum numbers
$J_\ell^{P_\ell} =0^-,1^-,2^-$. Their $I=1$ members are
$\Sigma_{c0}(\frac12^-), \Sigma_{c1}(\frac12^-,\frac32^-)$ and
$\Sigma_{c2}(\frac32^-,\frac52^-)$, while the corresponding
$I=\frac12$ members are $\Xi'_{c0}(\frac12^-),
\Xi'_{c1}(\frac12^-,\frac32^-)$ and
$\Xi'_{c2}(\frac32^-,\frac52^-)$ (see Table \ref{tab:pwave}), The
$J_\ell^{P_\ell}=0^-$ multiplet will be represented as a symmetric
matrix $(U)_{ij}$ defined in the same manner as $(B_6)_{ij}$
 \be \label{eq:U}
U_{ij}=\left( \begin{array}{ccc}
\Sigma_{c0}^{++} & \frac{1}{\sqrt2}\Sigma_{c0}^+ & \frac{1}{\sqrt2}\Xi'^+_{c0} \\
\frac{1}{\sqrt2}\Sigma_{c0}^+ & \Sigma_{c0}^0 & \frac{1}{\sqrt2}\Xi'^0_{c0} \\
\frac{1}{\sqrt2}\Xi'^+_{c0} & \frac{1}{\sqrt2}\Xi'^0_{c0} &
\Omega_{c0}^0 \end{array} \right)_{ij},
 \en
The $J_\ell^{P_\ell}=1^-$ multiplet will be represented as a
superfield similar to (\ref{eq:R}) but with a symmetric matrix
$\V_\mu^{ij}$
 \be \label{5.1}
\V_\mu^{ij} = \frac{1}{\sqrt3}(\gamma_\mu+v_\mu)\gamma_5 V^{ij} +
V^{*ij}_\mu,
 \en
where $V_{ij}$ has the same expression as $U_{ij}$ except for the
replacement of the superscript ``$c0$" by ``$c1$". The superfield
corresponding to the $J_\ell^{P_\ell}=2^-$ baryons is constructed
as \cite{Falk}
 \be
\X_{\mu\nu}^{ij} = X_{\mu\nu}^{*ij} +
\frac{1}{\sqrt{10}}\left\{(\gamma_\mu+v_\mu)\gamma_5
g_{\nu}^\alpha + (\gamma_\nu+v_\nu)\gamma_5
g_{\mu}^\alpha\right\}X_\alpha^{ij},
 \en
with $X_{\mu\nu}^{*ij}$ a spin-$\frac52$ Rarita-Schwinger field
and $X_\alpha^{ij}$ its spin-$\frac32$ heavy quark symmetry
partner.

The $p$-wave states with antisymmetric orbital wave functions can
be constructed in complete analogy to the symmetric ones.
Following \cite{Pirjol}, we use the superfield $\tilde
\R^{ij}_\mu$ constructed in analogy to $\S^{ij}_\mu$ to represent
the two sextets $\tilde\Sigma_{c1}(\frac12^-,\frac32^-)$ and
$\tilde\Xi'_{c1}(\frac12^-,\frac32^-)$. Likewsie, we use the
superfields $\tilde \U^i_\mu,\tilde \V^{i}_\mu,\tilde
\X^{i}_{\mu\nu}$ to denote the  antitriplets:
$\tilde\Lambda_{c0}^{+},\tilde\Lambda_{c1}^{+},\tilde\Lambda_{c2}^{+}$
in $I=0$ and $\tilde\Xi_{c0},\tilde\Xi_{c1},\tilde\Xi_{c2}$ in
$I=\frac12$.

The leading Lagrangian terms describing $P$-wave couplings among
the $s$-wave baryons  and $S$-wave couplings between the $s$-wave
and $p$-wave baryons are
 \be \label{eq:p}
{\cal L}_{P} = \frac32 ig_1
\epsilon_{\mu\nu\sigma\lambda}\mbox{Tr}(\bar \S^\mu v^\nu A^\sigma
\S^\lambda) - \sqrt3 g_2 \mbox{Tr}\left(\bar B_{\bar 3}A^\mu
\S_\mu + \bar \S^\mu A_\mu B_{\bar 3}\right),
 \en
and\footnote{The original $h_1$ term defined in \cite{Cho} is now
the $g_2$ term in Eq. (\ref{eq:p}) where we have followed
\cite{Yan} for the definition of $g_1$ and $g_2$ couplings. The
$h_3,\cdots, h_7$ terms were first introduced in \cite{Pirjol}.}
 \be \label{eq:s}
 {\cal L}_{S} &=& h_2\left\{ \epsilon_{ijk}\bar \R^{\mu}_i v_\nu A^\nu_{jl}\S_\mu^{kl}+
\epsilon_{ijk}\bar \S^{kl}_\mu v_\nu A^\nu_{lj} \R^{\mu}_i\right\}
+ h_3 \mbox{Tr}\left(\bar B_{\bar 3}v_\mu A^\mu U + \bar U v^\mu
A_\mu
B_{\bar 3}\right)\nonumber\\
&+& h_4\mbox{Tr}\left\{ \bar \V_\mu v_\nu A^\nu \S^\mu + \bar
\S_\mu v_\nu A^\nu \V^\mu\right\} + h_5\mbox{Tr}\left( \bar
{\tilde \R}_\mu
v_\nu A^\nu \S^\mu + \bar \S^\mu v_\nu A^\nu \tilde \R_\mu\right) \non \\
 &+&
h_6\left( \bar \T_i v_\nu A^\nu_{ji}\tilde \U_j + \bar {\tilde
\U}_i v_\nu A^\nu_{ji}\T_j\right) +h_7\left\{ \epsilon_{ijk}\bar
{\tilde \V}^{i}_\mu v_\nu A^\nu_{jl}\S^\mu_{kl} +
\epsilon_{ijk}\bar \S^\mu_{kl} v_\nu A^\nu_{lj}\tilde
\V^{i}_\mu\right\}\,,
 \en
respectively. The Goldstone bosons couple to the matter fields
through the nonlinear axial-vector field $A_\mu$ defined as
 \be A_\mu =
\frac{i}{2}\left( \xi^\dagger\partial_\mu\xi -
\xi\partial_\mu\xi^\dagger \right)=-{1\over
f_\pi}\partial_\mu\phi+{1\over
6f_\pi^3}[\phi,[\phi,\partial_\mu\phi]]+\cdots,
 \en
with $\xi=\exp(i\phi/f_\pi)$,
$\phi\equiv\frac{1}{\sqrt2}\pi^a\lambda^a$ and $f_\pi=132$ MeV.

The $D$-wave couplings of the $p$-wave baryons to $s$-wave baryons
are described by dimension-5 terms in the effective Lagrangian
\cite{Pirjol}
 \be  \label{eq:D} {\cal L}_{D} &=& ih_8\epsilon_{ijk}
\bar \S_\mu^{kl}\left( {\cal D}^\mu A^\nu + {\cal D}^\nu A^\mu
+ \frac23 g^{\mu\nu}(v\cdot {\cal D})(v\cdot A)\right)_{lj}\R_\nu^i \non \\
&+& ih_9 \mbox{Tr}\left\{\bar \S_\mu \left( {\cal D}^\mu A^\nu +
{\cal D}^\nu A^\mu
+ \frac23 g^{\mu\nu}(v\cdot {\cal D})(v\cdot A)\right) \V_\nu\right\} \non \\
&+& ih_{10} \epsilon_{ijk} \bar \T_i\left( {\cal D}_\mu A_\nu +
{\cal D}_\nu A_\mu\right)_{jl} \X_{kl}^{\mu\nu} +
h_{11}\epsilon_{\mu\nu\sigma\lambda}\mbox{Tr}\left\{ \bar
\S^{\mu}\left( {\cal D}^\nu A_\alpha + {\cal D}_\alpha
A^\nu\right)\X^{\alpha\sigma}
\right\}v^\lambda\nonumber\\
&+& ih_{12}\mbox{Tr}\left\{ \bar \S_\mu \left( {\cal D}^\mu A^\nu
+ {\cal D}^\nu A^\mu
+ \frac23 g^{\mu\nu}(v\cdot {\cal D})(v\cdot A)\right) \tilde \R_\nu\right\}\nonumber\\
&+& ih_{13}\epsilon_{ijk} \bar \S_\mu^{kl} \left( {\cal D}^\mu
A^\nu + {\cal D}^\nu A^\mu
+ \frac23 g^{\mu\nu}(v\cdot {\cal D})(v\cdot A)\right)_{lj} \tilde \V_\nu^{i}\nonumber\\
&+& ih_{14}\bar \T_i\left( {\cal D}^\mu A^\nu + {\cal D}^\nu
A^\mu\right)_{ji} \tilde \X_{\mu\nu}^{j} +
h_{15}\epsilon_{\mu\nu\sigma\lambda}\epsilon_{ijk} \bar \S^\mu_{
kl}\left( {\cal D}^\nu A_\alpha + {\cal D}_\alpha
A^\nu\right)_{lj} \tilde \X^{\alpha\sigma}_i v^\lambda \,,
 \en
where the covariant derivative of the axial-vector field $A_\mu$
is defined as ${\cal D}_\mu A_\nu = \partial_\mu A_\nu +
[V_\mu\,,A_\nu]$ with
 \be V_\mu = \frac{1}{2}\left(
\xi^\dagger\partial_\mu\xi + \xi\partial_\mu\xi^\dagger
\right)={1\over 2f_\pi^2}[\phi,\partial_\mu\phi]+\cdots\,,
 \en
and satisfies the relation ${\cal D}_\mu A_\nu - {\cal D}_\nu
A_\mu = 0$. Note that a pure $D$-wave is described by the
configuration
 \be
 D_{\mu\nu}=(\partial_\mu-v_\mu v\cdot \partial)(\partial_\nu-v_\nu v\cdot
 \partial)-\frac13(g_{\mu\nu}-v_\mu v_\nu)(\partial-v
 v\cdot\partial)^2,
 \en
satisfying $v^\mu D_{\mu\nu}=0$, $D_\mu^{~\mu}=0$ and
$D_{\mu\nu}=D_{\nu\mu}$. It is straightforward to show that the
structure ${\cal D}^\mu A^\nu + {\cal D}^\nu A^\mu + \frac23
g^{\mu\nu}(v\cdot {\cal D})(v\cdot A)$ appearing in Eq.
(\ref{eq:D}) indeed projects out a pure $D$-wave.

Some of the partial widths derived from the Lagrangians
(\ref{eq:p}) and (\ref{eq:s}) are \cite{Pirjol}:
 \begin{eqnarray} \label{eq:swavecoupling}
 \Gamma(\Sigma_c^*\to \Sigma_c\pi)={g_1^2\over 2\pi
 f_\pi^2}\,{m_{\Sigma_c}\over m_{\Sigma_c^*}}p_\pi^3, &&
 \Gamma(\Sigma_c\to \Lambda_c\pi)={g_2^2\over 2\pi
 f_\pi^2}\,{m_{\Lambda_c}\over m_{\Sigma_c}}p_\pi^3, \nonumber \\
 \Gamma(\Lambda_{c1}(1/2^-)\to \Sigma_c\pi)={h_2^2\over 2\pi
 f_\pi^2}\,{m_{\Sigma_c}\over m_{\Lambda_{c1}}}E_\pi^2p_\pi, &&
 \Gamma(\Sigma_{c0}(1/2^-)\to \Lambda_c\pi)={h_3^2\over 2\pi
 f_\pi^2}\,{m_{\Lambda_c}\over m_{\Sigma_{c0}}}E_\pi^2p_\pi,
 \nonumber \\
 \Gamma(\Sigma_{c1}(1/2^-)\to \Sigma_c\pi)={h_4^2\over 4\pi
 f_\pi^2}\,{m_{\Sigma_c}\over m_{\Sigma_{c1}}}E_\pi^2p_\pi, &&
 \Gamma(\tilde\Sigma_{c1}(1/2^-)\to \Sigma_c\pi)={h_5^2\over 4\pi
 f_\pi^2}\,{m_{\Sigma_c}\over m_{\tilde\Sigma_{c1}}}E_\pi^2p_\pi, \\
 \Gamma(\tilde\Xi_{c0}(1/2^-)\to \Xi_c\pi)={h_6^2\over 2\pi
 f_\pi^2}\,{m_{\Xi_c}\over m_{\tilde\Xi_{c0}}}E_\pi^2p_\pi, &&
 \Gamma(\tilde\Lambda_{c1}(1/2^-)\to \Sigma_c\pi)={h_7^2\over 2\pi
 f_\pi^2}\,{m_{\Sigma_c}\over m_{\tilde\Lambda_{c1}}}E_\pi^2p_\pi,
 \nonumber
 \end{eqnarray}
where $p_\pi$ is the c.m. momentum of the pion and $f_\pi=132$
MeV. Unfortunately, the decay $\Sigma_c^*\to\Sigma_c\pi$ is
kinematically prohibited since the mass difference between
$\Sigma_c^*$ and $\Sigma_c$ is only of order 65 MeV. Consequently,
the coupling $g_1$ cannot be extracted directly from the strong
decays of heavy baryons. Note that since the charge of the final
state is not specified in Eq. (\ref{eq:swavecoupling}), care must
be taken for the neutral pion state. For example, an additional
factor of $\frac12$ should be taken for
$\Sigma_c^{*+}\to\Sigma_c^+\pi^0$ and $\tilde
\Xi_{c0}^+\to\Xi_c^+\pi^0$, but not for
$\Sigma_c^+\to\Lambda_c^+\pi^0$.

In the quark model, various couplings in Eqs. (\ref{eq:s}) and
(\ref{eq:D}) are related to each other. The $S$-wave couplings
between the $s$-wave and the $p$-wave baryons are related by
\cite{Pirjol}
 \begin{eqnarray} \label{eq:QMh3}
 {|h_3|\over |h_4|}={\sqrt{3}\over 2}, \quad {|h_2|\over |h_4|}={1\over 2},
 \quad {|h_5|\over |h_6|}={2\over \sqrt{3}},\quad {|h_5|\over
 |h_7|}=1\,.
 \end{eqnarray}
The $D$-wave couplings satisfy the relations \cite{Pirjol}
 \begin{eqnarray} \label{eq:QMh8}
 |h_8|=|h_9|=|h_{10}|, \quad {|h_{11}|\over |h_{10}|}={|h_{15}|\over |h_{14}|}=\sqrt{2},
 \quad {|h_{12}|\over |h_{13}|}=2, \quad {|h_{14}|\over
 |h_{13}|}=1\,.
 \end{eqnarray}
From the dimensional analysis, it is expected that the dimensional
$D$-wave couplings $h_{8,\cdots,14}$ are of order
$1/\Lambda_\chi\sim (1.0-1.2)\times 10^{-3}\,{\rm MeV}^{-1}$,
where $\Lambda_\chi=(0.83\sim 1)$ GeV is the chiral symmetry
breaking scale.

As discussed in Sec. II, there exist first positive parity excited
charmed baryon states. Their orbital angular momentum $L$ is 2, 1
or 0. The transitions of these excited baryons to the $s$-wave
baryons involve $S$-wave, $P$-wave and $F$-wave couplings.
However, given the complications with the even parity excitations,
here we will not generalize HHChPT to include them.

In terms of the Wigner 6-j symbol, the decay rate for the baryon
decay process $J\to J'+\pi$ after spin-averaging over the initial
spin and summing over final spins has the expression
\cite{Peskin,Manohar}
 \be \label{eq:6j}
 \Gamma(J\to J'+\pi)=\,(2J_\ell+1)(2J'+1)\left|\left\{\matrix{L_\pi
& J'_\ell & J_\ell  \cr S_Q & J & J'\cr}\right\}\right|^2
p_\pi^{2L_\pi+1}|M_{L_\pi}|^2,
 \en
where $L_\pi$ is the orbital angular momentum of the pion,
$S_Q=\frac12$ is the heavy quark spin and $M_{L_\pi}$ is the
reduced matrix element which is independent of $J$ and $J'$. This
relation is very useful in relating strong decays into different
multiplets for a given partial wave.

\subsection{Strong decays of $s$-wave charmed baryons}
In the framework of heavy hadron chiral pertrubation theory
(HHChPT), one can use some measurements as input to fix the
coupling $g_2$ which, in turn, can be used to predict the rates of
other strong decays. We shall use the measured rates of
$\Sigma_c^{++}\to\Lambda_c^+\pi^+$,
$\Sigma_c^{*++}\to\Lambda_c^+\pi^+$ and
$\Sigma_c^{*0}\to\Lambda_c^+\pi^-$  as inputs to obtain
 \begin{eqnarray} \label{eq:g2}
 |g_2|=0.605^{+0.039}_{-0.043}\,,\qquad 0.57\pm0.04\,, \qquad
 0.60\pm0.04\,,
 \end{eqnarray}
respectively, where we have neglected the tiny contributions from
electromagnetic decays. Note that $|g_2|$ obtained from
$\Sigma_c^0\to\Lambda_c^+\pi^-$ has the same central value as the
first one in Eq. (\ref{eq:g2}) except that the errors are slightly
large.\footnote{Historically, based on the non-relativistic quark
model, the prediction $\Gamma(\Sigma_c^0\to\Lambda_c^+\pi^-)=2.45$
MeV was made long before experiment \cite{Yan}.}
Hence, the averaged $g_2$ is\footnote{For previous efforts of
extracting $g_2$ from experiment using HHChPT, see
\cite{Cheng97,Pirjol}. }
 \be
 |g_2|=0.591\pm0.023\,.
 \en

As pointed out in \cite{Yan}, within the framework of the
non-relativistic quark model, the couplings $g_1$ and $g_2$ can be
related to $g_A^q$, the axial-vector coupling in a single quark
transition of $u\to d$, via
 \begin{eqnarray}
 g_1={4\over 3}g_A^q, \qquad\qquad g_2=\sqrt{2\over 3}g_A^q.
 \end{eqnarray}
Using $g_A^q=0.75$ which is required to reproduce the correct
value of the nucleon axial coupling $g_A^N=1.25$, we obtain
  \begin{eqnarray}
 g_1=1, \qquad\qquad g_2=0.61\,.
 \end{eqnarray}
Hence, the quark model prediction is in good agreement with
experiment, while the large-$N_c$ prediction
$|g_2|=g_A^N/\sqrt{2}=0.88\,$ \cite{Guralnik} deviates from the
data by $2\sigma$. Applying (\ref{eq:g2}) leads to (see also Table
\ref{tab:strongdecayS})
 \begin{eqnarray}
\Gamma(\Xi_c^{*+})=\Gamma(\Xi_c^{*+}\to\Xi_c^+\pi^0,\Xi_c^0\pi^+)
&=& {g_2^2\over 4\pi
 f_\pi^2}\left({1\over 2}{m_{\Xi_c^+}\over m_{\Xi^{*+}_c}}p_\pi^3+
 {m_{\Xi_c^0}\over m_{\Xi^{*+}_c}}p_\pi^3\right)=
(2.7\pm 0.2)\,{\rm MeV},   \nonumber \\
\Gamma(\Xi_c^{*0})=\Gamma(\Xi_c^{*0}\to\Xi_c^+\pi^-,\Xi_c^0\pi^0)
&=& {g_2^2\over 4\pi
 f_\pi^2}\left({m_{\Xi_c^+}\over m_{\Xi^{*0}_c}}p_\pi^3+{1\over 2}
 {m_{\Xi_c^0}\over m_{\Xi^{*0}_c}}p_\pi^3\right)=
(2.8\pm 0.2)\,{\rm MeV}.
 \end{eqnarray}
Note that we have neglected the effect of $\Xi_c-\Xi'_c$ mixing in
calculations (for recent considerations, see \cite{Boyd,Ito}).
Therefore, the predicted total width of $\Xi_c^{*+}$ is in the
vicinity of the current limit $\Gamma(\Xi_c^{*+})<3.1$ MeV
\cite{CLEOb}.

\begin{table}[t]
\caption{Decay widths (in units of MeV) of $s$-wave charmed
baryons. Theoretical predictions of \cite{Tawfiq} are taken from
Table IV of \cite{Ivanov}.} \label{tab:strongdecayS}
\begin{center}
\begin{tabular}{|c|c|c|c|c|c|c|} \hline \hline
~~~~~~~Decay~~~~~~~ & Expt. & ~This work~ & Tawfiq & Ivanov &
Huang & Albertus
 \\
& ~~\cite{PDG}~~ & HHChPT & et al. \cite{Tawfiq} &
 et al. \cite{Ivanov} &  et al. \cite{Huang95} & ~~et al. \cite{Albertus}~~  \\
\hline
 $\Sigma_c^{++}\to\Lambda_c^+\pi^+$ & $2.23\pm0.30$ & input & $1.51\pm0.17$ & $2.85\pm0.19$
 &  2.5  & $2.41\pm0.07$  \\ \hline
 $\Sigma_c^{+}\to\Lambda_c^+\pi^0$ & $<4.6$ & $2.5\pm0.2$ & $1.56\pm0.17$ & $3.63\pm0.27$ &
 3.2 & $2.79\pm0.08$ \\ \hline
 $\Sigma_c^{0}\to\Lambda_c^+\pi^-$ & $2.2\pm0.4$ & input  & $1.44\pm0.16$ & $2.65\pm0.19$ &
 2.4 & $2.37\pm0.07$ \\ \hline
 $\Sigma_c(2520)^{++}\to\Lambda_c^+\pi^+$ & $14.9\pm1.9$ & input & $11.77\pm1.27$ & $21.99\pm0.87$ &
 8.2 & $17.52\pm0.75$   \\  \hline
 $\Sigma_c(2520)^{+}\to\Lambda_c^+\pi^0$ & $<17$ & $16.6\pm1.3$ & $$ & $$
 &8.6 & $17.31\pm0.74$  \\  \hline
 $\Sigma_c(2520)^{0}\to\Lambda_c^+\pi^-$ & $16.1\pm2.1$ & input & $11.37\pm1.22$ & $21.21\pm0.81$ &
 8.2 & $16.90\pm0.72$   \\  \hline
 $\Xi_c(2645)^+\to\Xi_c^{0,+}\pi^{+,0}$ & $<3.1$ & $2.7\pm0.2$ & $1.76\pm0.14$ & $3.04\pm0.37$ &
 & $3.18\pm0.10$  \\  \hline
 $\Xi_c(2645)^0\to\Xi_c^{+,0}\pi^{-,0}$ & $<5.5$ & $2.8\pm0.2$ & $1.83\pm0.06$ & $3.12\pm0.33$ &
 & $3.03\pm0.10$  \\ \hline \hline

\end{tabular}
\end{center}
\end{table}

It is clear from Table \ref{tab:strongdecayS} that  the strong
decay width of $\Sigma_c$ is smaller than that of $\Sigma_c^*$ by
a factor of $\sim 7$, although they will become the same in the
limit of heavy quark symmetry. This is ascribed to the fact that
the c.m. momentum of the pion is around 90 MeV in the decay
$\Sigma_c\to\Lambda_c\pi$ while it is two times bigger in
$\Sigma_c^*\to\Lambda_c\pi$. Since $\Sigma_c$ states are
significantly narrower than their spin-$\frac32$ counterparts,
this explains why the measurement of their widths came out much
later. Instead of using the data to fix the coupling constants in
a model-independent manner, there exist some calculations of
couplings in various models such as the relativistic light-front
model \cite{Tawfiq}, the relativistic three-quark model
\cite{Ivanov} and light-cone sum rules \cite{Huang95,Zhu}. The
calculated results are summarized in Table \ref{tab:strongdecayS}.

It is worth remarking that although the coupling $g_1$ cannot be
determined directly from the strong decay such as
$\Sigma_c^*\to\Sigma_c\pi$, some information of $g_1$ can be
learned from the radiative decay $\Xi_c^{*0}\to\Xi_c^0 \gamma$,
which is prohibited at tree level by SU(3) symmetry but can be
induced by chiral loops. A measurement of
$\Gamma(\Xi_c^{*0}\to\Xi_c^0\gamma)$ will yield two possible
solutions for $g_1$.  Assuming the validity of the quark model
relations among different coupling constants, the experimental
value of $g_2$ implies $|g_1|=0.93\pm 0.16$ \cite{Cheng97}.

\subsection{Strong decays of $p$-wave charmed baryons}

Some of the $S$-wave and $D$-wave couplings of $p$-wave baryons to
$s$-wave baryons can be determined. In principle, the coupling
$h_2$ is readily extracted from $\Lambda_c(2593)^+\to
\Sigma_c^0\pi^+$ with $\Lambda_c(2593)$ being identified as
$\Lambda_{c1}(\frac12^-)$. However, since
$\Lambda_c(2593)^+\to\Sigma_c\pi$ is kinematically barely allowed,
the finite width effects of the intermediate resonant states could
become important \cite{Falk03}. If these effects are neglected,
then from Eq. (\ref{eq:swavecoupling}) and the measured decay
rates of $\Lambda_c(2593)^+\to\Sigma_c^0\pi^+$ and
$\Lambda_c(2593)^+\to\Sigma_c^{++}\pi^-$, we find
 \begin{eqnarray} \label{eq:h2}
|h_2|=0.41\pm0.11\,.
 \end{eqnarray}

Before proceeding to a more precise determination of $h_2$, we
make several remarks on the partial widths of $\Lambda_c(2593)^+$
decays. (i) PDG \cite{PDG} has assumed the isospin relation,
namely,
$\Gamma(\Lambda_c^+\pi^+\pi^-)=2\Gamma(\Lambda_c^+\pi^0\pi^0)$ to
extract the branching ratios for $\Sigma_c\pi$ modes. However, the
decay $\Lambda_c(2593)\to\Lambda_c\pi\pi$ occurs very close to the
threshold as $m_{\Lambda_c(2593)}-m_{\Lambda_c}=308.9\pm0.6$ MeV.
Hence, the phase space is very sensitive to the small
isospin-violating mass differences between members of pions and
charmed Sigma baryon multiplets. Since the neutral pion is
slightly lighter than the charged one, it turns out that both
$\Lambda_c^+\pi^+\pi^-$ and $\Lambda_c^+\pi^0\pi^0$ have very
similar rates [see Eq. (\ref{eq:3body}) below]. (ii) Taking
$\B(\Lambda_c(2593)^+\to \Lambda_c^+\pi^+\pi^-)\approx 0.5$ and
using the measured ratios \cite{PDG}
 \be
 {\Gamma(\Lambda_c(2593)^+\to \Sigma_c^{++}\pi^-)\over
\Gamma(\Lambda_c(2593)^+\to \Lambda_c^+\pi^+\pi^-)}=0.36\pm0.10,
\qquad  {\Gamma(\Lambda_c(2593)^+\to \Sigma_c^{0}\pi^+)\over
\Gamma(\Lambda_c(2593)^+\to \Lambda_c^+\pi^+\pi^-)}=0.37\pm0.10,
 \en
we obtain
 \be
 \B(\Lambda_c(2593)^+\to\Sigma_c^{++}\pi^-)=0.18\pm0.05, \qquad
 \B(\Lambda_c(2593)^+\to\Sigma^{0}_c\pi^+)=0.19\pm0.05\,,
 \en
and
 \be
 \Gamma(\Lambda_c(2593)^+\to\Sigma_c^{++}\pi^-)=0.65^{+0.41}_{-0.31}\,{\rm MeV}, \qquad
 \Gamma(\Lambda_c(2593)^+\to\Sigma_c^{0}\pi^+)=0.67^{+0.41}_{-0.31}\,{\rm
 MeV}\,.
 \en
(iii) The non-resonant or direct three-body decay mode
$\Lambda_c^+\pi^+\pi^-$ has a branching ratio of $0.14\pm0.08$
\cite{PDG}. Assuming the same for $\Lambda_c^+\pi^0\pi^0$, then
the fractions for resonant and non-resonant $\Lambda_c^+\pi\pi$
are $0.73 \pm0.15$ and $0.27\pm0.15$, respectively, where
$\Lambda_c^+\pi\pi=\Lambda_c^+\pi^+\pi^-+\Lambda_c^+\pi^0\pi^0$.
From the measured total width of $\Lambda_c(2593)^+$,
$3.6^{+2.0}_{-1.3}$ MeV \cite{PDG}, we are led to
 \be
 \Gamma(\Lambda_c(2593)^+\to\Lambda_c^+\pi\pi)_{\rm R} &=& (2.63^{+1.56}_{-1.09})\,{\rm
 MeV}, \non \\
 \Gamma(\Lambda_c(2593)^+\to\Lambda_c^+\pi\pi)_{\rm NR} &=&
 (0.97^{+0.76}_{-0.64})\,{\rm
 MeV}.
 \en

\begin{table}[t]
\caption{Same as Table \ref{tab:strongdecayS} except for $p$-wave
charmed baryons. } \label{tab:strongdecayP}
\begin{center}
\begin{tabular}{|c|c|c|c|c|c|c|} \hline \hline
~~~~~~~Decay~~~~~~~ & Expt. & This work & Tawfiq & Ivanov & Huang & Zhu   \\
& ~~\cite{PDG}~~& ~HHChPT~ & ~~et al. \cite{Tawfiq}~~ &
 ~~~et al. \cite{Ivanov}~~~ &  ~~et al. \cite{Huang95}~~ & ~~\cite{Zhu}~~  \\
\hline
 $\Lambda_c(2593)^+\to (\Lambda_c^{+}\pi\pi)_R$ & $2.63^{+1.56}_{-1.09}$ & input & $$
 & $$ & $2.5$ &  \\ \hline
 $\Lambda_c(2593)^+\to \Sigma_c^{++}\pi^-$ & $0.65^{+0.41}_{-0.31}$ & $0.72^{+0.43}_{-0.30}$ & $1.47\pm0.57$ &
 $0.79\pm0.09$ & $0.55^{+1.3}_{-0.55}$ & 0.64  \\ \hline
 $\Lambda_c(2593)^+\to \Sigma_c^{0}\pi^+$ & $0.67^{+0.41}_{-0.31}$ & $0.77^{+0.46}_{-0.32}$ & $1.78\pm0.70$
 & $0.83\pm0.09$ & $0.89\pm0.86$ & 0.86 \\ \hline
 $\Lambda_c(2593)^+\to \Sigma_c^{+}\pi^0$ & & $1.57^{+0.93}_{-0.65}$ & $1.18\pm0.46$
 & $0.98\pm0.12$ & $1.7\pm0.49$ & 1.2  \\ \hline
 $\Lambda_c(2625)^+\to \Sigma_c^{++}\pi^-$ & $<0.10$ & $\lsim 0.029$ & $0.44\pm0.23$ & $0.076\pm0.009$ &
 $0.013$ & 0.011  \\ \hline
 $\Lambda_c(2625)^+\to \Sigma_c^{0}\pi^+$ & $<0.09$ & $\lsim 0.029$ & $0.47\pm0.25$ & $0.080\pm0.009$
 & 0.013 & 0.011  \\ \hline
 $\Lambda_c(2625)^+\to \Sigma_c^{+}\pi^0$ & & $\lsim 0.041$ & $0.42\pm0.22$ & $0.095\pm0.012$
 & 0.013 & 0.011  \\ \hline
 $\Lambda_c(2625)^+\to \Lambda_c^+\pi\pi$ & $<1.9$ & $\lsim 0.21$ & $$ & $$
 & 0.11 &   \\ \hline
 $\Sigma_c(2800)^{++}\to\Lambda_c\pi,\Sigma_c^{(*)}\pi$ & $75^{+22}_{-17}$ & input &
 & & &  \\ \hline
 $\Sigma_c(2800)^{+}\to\Lambda_c\pi,\Sigma_c^{(*)}\pi$ & $62^{+60}_{-40}$ & input &
 & & &  \\ \hline
 $\Sigma_c(2800)^0\to\Lambda_c\pi,\Sigma_c^{(*)}\pi$ & $61^{+28}_{-18}$ & input &
 & & & \\ \hline
 $\Xi_c(2790)^+\to\Xi'^{0,+}_c\pi^{+,0}$ & $<15$ & $8.0^{+4.7}_{-3.3}$ &
 $$ & $$ & &  \\ \hline
 $\Xi_c(2790)^0\to\Xi'^{+,0}_c\pi^{-,0}$  & $<12$ & $8.5^{+5.0}_{-3.5}$ &
 $$ & $$ & & \\ \hline
 $\Xi_c(2815)^+\to\Xi^{*+,0}_c\pi^{0,+}$ & $<3.5$ & $3.4^{+2.0}_{-1.4}$ &
 $2.35\pm0.93$ & $0.70\pm0.04$ & &  \\ \hline
 $\Xi_c(2815)^0\to\Xi^{*+,0}_c\pi^{-,0}$ & $<6.5$ & $3.6^{+2.1}_{-1.5}$ &
 $$ & $$ & &  \\ \hline \hline
\end{tabular}
\end{center}
\end{table}

Pole contributions to the decays
$\Lambda_c(2593)^+,\Lambda_c(2625)^+\to \Lambda_c^+\pi\pi$ have
been considered in \cite{Cho,Huang95,Pirjol} with the finite width
effects included. The intermediate states of interest are
$\Sigma_c$ and $\Sigma_c^*$ poles. The resonant contribution
arises from the $\Sigma_c$ pole, while the non-resonant term
receives a contribution from the $\Sigma_c^*$ pole. (Since
$\Lambda_c(2593)^+,\Lambda_c(2625)^+\to\Sigma_c^*\pi$ are not
kinematically allowed, the $\Sigma_c^*$ pole is not a resonant
contribution.) The decay rates thus depend on two coupling
constants $h_2$ and $h_8$. The decay rate for the process
$\Lambda_{c_1}^+(2593)\to \Lambda_c^+\pi^+\pi^-$ can be calculated
in the framework of heavy hadron chiral perturbation theory
\cite{Pirjol}
 \be
& & {d^2\Gamma(\Lambda_{c1}^{+}(2593)\to
\Lambda_c^+\pi^+(E_1)\pi^-(E_2))\over dE_1dE_2}=  \\
& &\qquad \frac{g_2^2}{16\pi^3 f_\pi^4}m_{\Lambda_c^+}\left\{ {\bf
p}_2^2|A|^2 + {\bf p}_1^2|B|^2 + 2{\bf p}_1\cdot{\bf p}_2\,
\mbox{Re }(AB^*)\right\},  \nonumber
 \en
with
 \be
 && A(E_1,E_2)
 = \frac{h_2E_1}{\Delta_R-\Delta_{\Sigma_c^0}-E_1+
i\Gamma_{\Sigma_c^0}/2}-\frac{\frac23 h_8{\bf p}_2^2}
{\Delta_R-\Delta_{\Sigma_c^{*0}}-E_1+i\Gamma_{\Sigma_c^{*0}}/2}, \\
& & \qquad\qquad  +\frac{2h_8{\bf p}_1\cdot{\bf p}_2}
{\Delta_R-\Delta_{\Sigma_c^{*++}}-E_2+i\Gamma_{\Sigma_c^{*++}}/2}\,,
\nonumber\\
&& B(E_1,E_2;\Delta_{\Sigma_c^{(*)0}},\Delta_{\Sigma_c^{(*)++}}) =
A(E_2,E_1;\Delta_{\Sigma_c^{(*)++}},\Delta_{\Sigma_c^{(*)0}})\,,
 \en
where $\Delta _R=m_{\Lambda_c(2593)}-m_{\Lambda_c}$ and $\Delta
_{\Sigma_c^{(*)}}=m_{\Sigma_c^{(*)}}-m_{\Lambda_c}$.

For the spin-${3\over 2}$ state $\Lambda_c(2625)$, its decay is
dominated by the three-body channel $\Lambda_c^+\pi\pi$ as the
major two-body decay $\Sigma_c\pi$ is a $D$-wave one. As for the
decay $\Lambda_{c1}^{+}(2625)\to \Lambda_c^+\pi^+\pi^-$, its rate
is given by \cite{Pirjol}
 \be
& & {d^2\Gamma(\Lambda_{c1}^{+}(2625)\to
\Lambda_c^+\pi^+(E_1)\pi^-(E_2))\over dE_1dE_2} =  \\
& &\quad \frac{g_2^2}{16\pi^3 f_\pi^4}M_{\Lambda_c^+}\left\{ {\bf
p}_1^2|C|^2 + {\bf p}_2^2|E|^2 + 2{\bf p}_1\cdot {\bf p}_2
\mbox{Re }(CE^*) +
[{\bf p}_1^2{\bf p}_2^2-({\bf p}_1\cdot {\bf p}_2)^2]\right.\times \nonumber\\
& &\left.\quad\,\,\,\, \left[ {\bf p}_1^2|D|^2 + {\bf p}_2^2|F|^2
- \mbox{Re }(CF^*) + \mbox{Re }(DE^*) + 2{\bf p}_1\cdot{\bf p}_2\,
\mbox{Re }(DF^*)\right]\right\}, \nonumber
 \en
with
 \be
& &C(E_1,E_2) = \left(h_2E_2-\frac23 h_8{\bf p}_2^2\right)
\frac{1}{\Delta_{R^*}-\Delta_{\Sigma_c^{*++}}-E_2+
i\Gamma_{\Sigma_c^{*++}}/2} \non \\
& &\qquad+\,\frac23 h_8\,{\bf p}_1\cdot {\bf p}_2
\left(\frac{1}{\Delta_{R^*}-\Delta_{\Sigma_c^{0}}-E_1+
i\Gamma_{\Sigma_c^{0}}/2} +
\frac{2}{\Delta_{R^*}-\Delta_{\Sigma_c^{*0}}-E_1+
i\Gamma_{\Sigma_c^{*0}}/2}\right), \non \\
& &D(E_1,E_2) = \frac23 h_8\left(
-\frac{1}{\Delta_{R^*}-\Delta_{\Sigma_c^{0}}-E_1+
i\Gamma_{\Sigma_c^{0}}/2} +
\frac{1}{\Delta_{R^*}-\Delta_{\Sigma_c^{*0}}-E_1+
i\Gamma_{\Sigma_c^{*0}}/2}\right),   \non \\
& &E(E_1,E_2;\Delta_{\Sigma_c^{(*)0}},\Delta_{\Sigma_c^{(*)++}}) =
C(E_2,E_1;\Delta_{\Sigma_c^{(*)++}},\Delta_{\Sigma_c^{(*)0}})\,, \non \\
& &F(E_1,E_2;\Delta_{\Sigma_c^{(*)0}},\Delta_{\Sigma_c^{(*)++}}) =
-D(E_2,E_1;\Delta_{\Sigma_c^{(*)++}},\Delta_{\Sigma_c^{(*)0}})\,,
 \en
where $\Delta _{R^*}=m_{\Lambda_c(2625)}-m_{\Lambda_c}$.

The total widths of the $\Lambda_c(2593)$ and $\Lambda_c(2625)$
states obtained after integrating out the variables $E_1$ and
$E_2$ and including the $\pi^0\pi^0$ channel are
 \be \label{eq:h2h8}
 \Gamma(\Lambda_c(2593)^+\to\Lambda_c^+\pi\pi)&=& 13.82h_2^2
 +26.28h_8^2-2.97h_2h_8, \non \\
 \Gamma(\Lambda_c(2625)^+\to\Lambda_c^+\pi\pi)&=& 0.617h_2^2+0.136\times
 10^6h_8^2-27h_2h_8,
 \en
where use of (\ref{eq:g2}) for $g_2$ has been made. It is clear
that the experimental limit on $\Gamma(\Lambda_c(2625))$ gives an
upper bound on $h_8$ of order $10^{-3}$ (in units of MeV$^{-1}$),
whereas the decay width of $\Lambda_c(2593)$ is entirely governed
by the coupling $h_2$. This indicates that the direct non-resonant
$\Lambda_c^+\pi\pi$ contribution cannot be described by the
$\Sigma_c^*$ pole alone. Some other mechanisms are needed to
account for the non-resonant contributions. Identifying the
calculated $\Gamma(\Lambda_c(2593)^+\to\Lambda_c^+\pi\pi)$ with
the resonant one, we find
 \be \label{eq:h2fw}
 |h_2|=0.437^{+0.114}_{-0.102}\,, \qquad\quad |h_8|< 3.65\times
 10^{-3}\,.
 \en
Comparing (\ref{eq:h2fw}) with (\ref{eq:h2}) we see that the
magnitude of $h_2$ is enhanced slightly by finite width effects.

Assuming that the total width of $\Lambda_c(2593)^+$ is saturated
by the resonant $\Lambda_c^+\pi\pi$ 3-body decays, Pirjol and Yan
obtained $|h_2|=0.572^{+0.322}_{-0.197}$ and $|h_8|\leq
(3.50-3.68)\times 10^{-3}\,{\rm MeV}^{-1}$ \cite{Pirjol}. Using
the updated hadron masses and
$\Gamma(\Lambda_c(2593))$,\footnote{The CLEO result
$\Gamma(\Lambda_c(2593))=3.9^{+2.4}_{-1.6}$ MeV \cite{CLEO} is
used in \cite{Pirjol} to fix $h_2$.}
we find $|h_2|=0.499^{+0.134}_{-0.100}$. Taking into account the
fact that the $\Sigma_c$ and $\Sigma_c^*$ poles only describe the
resonant contributions to the total width of $\Lambda_c(2593)$, we
finally reach at the value of $h_2$  given by (\ref{eq:h2fw}).
Using this result for $h_2$, the two-body
$\Lambda_c(2593)\to\Sigma_c\pi$ rates are shown in Table
\ref{tab:strongdecayP}. The three-body partial rates are found to
be
 \be \label{eq:3body}
 \Gamma(\Lambda_c(2593)^+\to\Lambda_c^+\pi^+\pi^-)=1.29\,{\rm MeV},\qquad
 \Gamma(\Lambda_c(2593)^+\to\Lambda_c^+\pi^0\pi^0)=1.34\,{\rm MeV}.
 \en
Therefore, isospin violation is manifested in the relations
$\Gamma(\Sigma_c^+\pi^0)\approx
2\Gamma(\Sigma_c^{++}\pi^-)\sim2\Gamma(\Sigma_c^{0}\pi^+)$ and
$\Gamma(\Lambda_c^+\pi^0\pi^0)\approx
\Gamma(\Lambda_c^+\pi^+\pi^-)$ in $\Lambda_c(2593)$ decays.

The $\Xi_c(2790)$ and $\Xi_c(2815)$ baryons form a doublet
$\Xi_{c1}({1\over 2}^-,{3\over 2}^-)$. $\Xi_c(2790)$ decays to
$\Xi'_c\pi$, while $\Xi_c(2815)$ decays to $\Xi_c\pi\pi$,
resonating through $\Xi^*_c$, i.e. $\Xi_c(2645)$. Using the
coupling $h_2$ obtained from (\ref{eq:h2fw}) and the experimental
observation that the $\Xi_c\pi\pi$ mode in $\Xi_c(2815)$ decays is
consistent with being entirely via $\Xi^*_c\pi$
\cite{CLEO:Xic2815}, the predicted $\Xi_c(2790)$ and $\Xi_c(2815)$
widths are shown in Table \ref{tab:strongdecayP}, where uses have
been made of
 \begin{eqnarray}
\Gamma(\Xi_{c1}(1/2)^+)\approx\Gamma(\Xi_{c1}(1/2)^+\to\Xi'^+_c\pi^0,\Xi'^0_c\pi^+)
&=& {h_2^2\over 4\pi
 f_\pi^2}\left({1\over 2}{m_{\Xi'^+_c}\over m_{\Xi_{c1}(1/2)}}E_\pi^2p_\pi+
 {m_{\Xi'^0_c}\over m_{\Xi_{c1}(1/2)}}E_\pi^2p_\pi\right),   \nonumber \\
\Gamma(\Xi_{c1}(3/2)^+)\approx\Gamma(\Xi_{c1}(3/2)^+\to\Xi^{*+}_c\pi^0,\Xi^{*0}_c\pi^+)
&=& {h_2^2\over 4\pi
 f_\pi^2}\left(\frac12{m_{\Xi^{*+}_c}\over m_{\Xi_{c1}(3/2)}}E_\pi^2p_\pi+
 {m_{\Xi^{*0}_c}\over m_{\Xi_{c1}(3/2)}}E_\pi^2p_\pi\right), \non\\
 \end{eqnarray}
and similar expressions for the neutral
$\Xi_{c1}(\frac12^-,\frac32^-)$ states. The predictions are
consistent with the current experimental limits.

Some information on the coupling $h_{10}$ can be inferred from the
strong decays of $\Sigma_c(2800)$. As noticed in passing, the
states $\Sigma_c(2800)^{++,+,0}$ which are observed in the
$\Lambda_c^+\pi$ spectrum are most likely to be
$\Sigma_{c2}({3\over 2}^-)$. From Table \ref{tab:pwave} we see
that there are three low-lying $p$-wave $\Sigma_c$ multiplets:
$\Sigma_{c0}$, $\Sigma_{c1}$ and $\Sigma_{c2}$. Both $\Sigma_{c0}$
and $\Sigma_{c2}$  decay to $\Lambda_c\pi$ in an $S$-wave and a
$D$-wave, respectively, while $\Sigma_{c1}$ decays mainly to the
two pion system $\Lambda_c\pi\pi$ in a $P$-wave. From Eqs.
(\ref{eq:swavecoupling}), (\ref{eq:h2fw}) and (\ref{eq:QMh3}) we
find $\Gamma(\Sigma_{c0}\to\Lambda_c\pi)\approx 406$ MeV. Hence,
it is too broad to be observable. Therefore,
$\Sigma_c(2800)^{++,+,0}$ are likely to be $\Sigma_{c2}({3\over
2}^-)$. Assuming their widths are dominated by the two-body
$D$-wave modes $\Lambda_c\pi$, $\Sigma_c\pi$ and $\Sigma_c^*\pi$,
we have \cite{Pirjol}
 \begin{eqnarray}
\Gamma\left(\Sigma_{c2}({3/2})^{++}\right) &\approx&
\Gamma\left(\Sigma_{c2}({3/ 2})^{++}\to\Lambda_c^+\pi^+\right)
\non \\ &+&
\Gamma\left(\Sigma_{c2}({3/2})^{++}\to\Sigma_c^+\pi^+\right)+
\Gamma\left(\Sigma_{c2}({3/ 2})^{++}\to\Sigma_c^{*+}\pi^+\right)
\non \\ &+&
\Gamma\left(\Sigma_{c2}({3/2})^{++}\to\Sigma_c^{++}\pi^0\right)+
\Gamma\left(\Sigma_{c2}({3/ 2})^{++}\to\Sigma_c^{*++}\pi^0\right),
 \en
and similar expressions for $\Sigma_{c2}(\frac32)^+$ and
$\Sigma_{c2}(\frac32)^0$. Using\footnote{It is useful to apply Eq.
(\ref{eq:6j}) to check the consistency of the partial decay rate
formulas.}
 \be
\Gamma\left(\Sigma_{c2}({3/ 2}^-)\to\Lambda_c\pi\right)
 &=& {4h_{10}^2\over 15\pi f_\pi^2}\,{m_{\Lambda_c}\over
m_{\Sigma_{c2}}}p_\pi^5, \non \\
\Gamma\left(\Sigma_{c2}({3/2}^-)\to\Sigma_c^{(*)}\pi\right) &=&
{h_{11}^2\over 10\pi f_\pi^2}\,{m_{\Sigma_c^{(*)}}\over
m_{\Sigma_{c2}}}p_\pi^5,
 \end{eqnarray}
and the quark model relation $h_{11}^2=2h_{10}^2$ [cf. Eq.
(\ref{eq:QMh8})] and the measured widths of
$\Sigma_c(2800)^{++,+,0}$ (Table \ref{tab:spectrum}), we obtain
 \begin{eqnarray}
|h_{10}|=(0.86^{+0.08}_{-0.10})\times 10^{-3}\,{\rm MeV}^{-1}\,.
 \end{eqnarray}
This is consistent with the naive expectation that $h_{10}\sim
1/\Lambda_\chi$. Since the state $\Sigma_{c1}({3\over 2}^-)$ is
broader, even a small mixing of $\Sigma_{c2}({3\over 2}^-)$ with
$\Sigma_{c1}({3\over 2}^-)$ could enhance the decay width of the
former \cite{Pirjol}. Moreover, the non-resonant three-body mode
$\Lambda_c^+\pi\pi$ may have contributions to the width of
$\Sigma_{c2}$,  the above value for $h_{10}$ should be regarded as
an upper limit of $|h_{10}|$ . Using the quark model relation
$|h_8|=|h_{10}|$ [Eq. (\ref{eq:QMh8})], we then have
 \begin{eqnarray} \label{eq:h8}
|h_8|\lsim (0.86^{+0.08}_{-0.10})\times 10^{-3}\,{\rm MeV}^{-1}\,,
 \end{eqnarray}
which improves the previous limit (\ref{eq:h2fw}) by a factor of
4.

Using the above value of $h_8$, the rates of $\Lambda_c(2625)$
decays to $\Lambda_c\pi\pi$ and $\Sigma_c\pi$ are presented in
Table \ref{tab:strongdecayP}, where we have used Eq.
(\ref{eq:h2h8}) and
 \be
 \Gamma(\Lambda_{c1}(3/2^-)\to\Sigma_c\pi)={2h_8^2\over 9\pi
 f_\pi^2}\,{m_{\Sigma_c}\over m_{\Lambda_{c1}(3/2)}}\,p_\pi^5.
 \en

\subsection{Strong decays of first positive parity excited charmed baryons}

Besides the $p$-wave charmed baryons discussed in the previous
subsection, some of the higher orbitally excited charmed baryons
listed in Table \ref{tab:newbaryons} are likely to be the first
positive parity excitations. For example, the recent Belle studies
favor the $J^P$ quantum numbers of $\Lambda_c(2880)^+$  to be
$\frac52^+$ \cite{Belle:Lamc2880}. The quantum numbers for the
first positive parity excited charmed baryons are listed in Table
\ref{tab:pp}. Those states have $L_K+L_k=2$ and hence the orbital
angular momentum $L_\ell$ can be 2, 1 or 0. Besides
$\Lambda_c(2880)^+$, the states $\Lambda_c(2765)^+$,
$\Lambda_c(2940)^+$, $\Xi_c(2980)^{+,0}$ and $\Xi_c(3077)^{+,0}$
are also likely to be the first positive parity excitations of
charmed baryons as we are going to discuss.

As noticed in Sec. II, Belle has studied the experimental
constraint on the $J^P$ quantum numbers of $\Lambda_c(2880)^+$ and
found that the assignment of $J=\frac52$ is favored over
$J=\frac12$ or $\frac32$ by the angular analysis of
$\Lambda_c(2880)^+\to\Sigma_c^{0,++}\pi^\pm$
\cite{Belle:Lamc2880}. The measurement of the ratio of
$\Lambda_c(2880)$ partial widths \cite{Belle:Lamc2880}
 \be \label{eq:r}
R={\Gamma(\Lambda_c(2880)\to \Sigma_c^*\pi^\pm)\over
\Gamma(\Lambda_c(2880)\to\Sigma_c\pi^\pm)}=(24.1\pm6.4^{+1.1}_{-4.5})\%
 \en
can be used to determine the parity assignment. From Tables
\ref{tab:pwave} and \ref{tab:pp} we see that the candidates for
the spin-${5\over 2}$ state are $\tilde\Lambda_{c2}(\frac52^-)$,
$\Lambda_{c2}(\frac52^+)$, $\tilde\Lambda'_{c2}(\frac52^+)$,
$\tilde\Lambda''_{c2}(\frac52^+)$ and
$\tilde\Lambda''_{c3}(\frac52^+)$. For $J^P=\frac52^-$,
$\Lambda_c(2880)$ decays to $\Sigma_c^*\pi$ and $\Sigma_c\pi$ in a
$D$ wave. From Eq. (\ref{eq:6j}) we obtain
 \be
 {\Gamma\left(\tilde\Lambda_{c2}(5/2^-)\to
[\Sigma_c^*\pi]_D\right)\over
\Gamma\left(\tilde\Lambda_{c2}(5/2^-)\to
[\Sigma_c\pi]_D\right)}={7\over 2}
\,{p_\pi^5(\Lambda_c(2880)\to\Sigma_c^*\pi)\over
p_\pi^5(\Lambda_c(2880)\to\Sigma_c\pi)}={7\over 2}\times
0.42=1.45\,.
 \en
Hence, the assignment of $J^P={5\over 2}^-$ for $\Lambda_c(2880)$
is disfavored. For $J^P=\frac52^+$, $\Lambda_{c2}$,
$\tilde\Lambda'_{c2}$ and $\tilde\Lambda''_{c2}$  with $J_\ell=2$
decay to $\Sigma_c\pi$ in a $F$ wave and $\Sigma_c^*\pi$ in $F$
and $P$ waves. Neglecting the $P$-wave contribution for the
moment,
 \be \label{eq:0.23}
 {\Gamma\left(\Lambda_{c2}(5/2^+)\to
[\Sigma_c^*\pi]_F\right)\over \Gamma\left(\Lambda_{c2}(5/2^+)\to
[\Sigma_c\pi]_F\right)}={4\over
5}\,{p_\pi^7(\Lambda_c(2880)\to\Sigma_c^*\pi)\over
p_\pi^7(\Lambda_c(2880)\to\Sigma_c\pi)}={4\over 5}\times
0.29=0.23\,.
 \en
At first glance, it appears that this is in good agreement with
experiment. However, the $\Sigma_c^*\pi$ channel is available via
a $P$-wave and is enhanced by a factor of $1/p_\pi^4$ (or more
precisely, $(\Lambda_\chi/p_\pi)^4$) relative to the $F$-wave one.
Unfortunately, we cannot apply Eq. (\ref{eq:6j}) to calculate the
contribution of the $[\Sigma_c^*\pi]_F$ channel to the ratio $R$
as the reduced matrix elements are different for $P$-wave and
$F$-wave modes. In any event, the $\Sigma_c^*\pi$ mode produced in
$\Lambda_c(2880)$ is {\it a priori} not necessarily suppressed
relative to $[\Sigma_c\pi]_F$. Therefore, if $\Lambda_c(2880)^+$
is one of the states $\Lambda_{c2}$, $\tilde\Lambda'_{c2}$ and
$\tilde\Lambda''_{c2}$, the prediction $R=0.23$ is not robust as
it can be easily upset by the contribution from the $P$-wave
$\Sigma_c^*\pi$.

As for $\tilde\Lambda''_{c3}(\frac52^+)$, it decays to
$\Sigma_c^*\pi$, $\Sigma_c\pi$ and $\Lambda_c\pi$ all in $F$
waves. Since $J_\ell=3,L_\ell=2$, it turns out that
 \be
 {\Gamma\left(\Lambda''_{c3}(5/2^+)\to
[\Sigma_c^*\pi]_F\right)\over \Gamma\left(\Lambda''_{c3}(5/2^+)\to
[\Sigma_c\pi]_F\right)}={5\over
4}\,{p_\pi^7(\Lambda_c(2880)\to\Sigma_c^*\pi)\over
p_\pi^7(\Lambda_c(2880)\to\Sigma_c\pi)}={5\over 4}\times
0.29=0.36\,.
 \en
Although this deviates from the experimental measurement
(\ref{eq:r}) by $1\sigma$, it is a robust prediction. However,
there are two issues with this assignment. First,
$\Lambda''_{c3}(\frac52^+)$ can decay to a $F$-wave $\Lambda_c\pi$
and this has not been seen by BaBar and Belle. Second, the quark
model indicates a $\Lambda_{c2}(\frac52^+)$  state around 2910 MeV
which is close to the mass of $\Lambda_c(2880)$, while the mass of
$\Lambda''_{c3}(\frac52^+)$ is higher \cite{Capstick}. Therefore,
we conjecture that the first positive-parity excited charmed
baryon $\Lambda_c(2880)^+$ could be an admixture of
$\Lambda_{c2}(\frac52^+)$ and $\Lambda''_{c3}(\frac52^+)$.

The quark potential model predicts a $\frac52^-$ $\Lambda_c$ state
at 2900 MeV and a $\frac32^+$ $\Lambda_c$ state at 2910 MeV
\cite{Capstick}. Given the uncertainty of order 50 MeV for the
quark model calculation, this suggests that the possible allowed
$J^P$ numebrs of the highest $\Lambda_c(2940)^+$ are $\frac52^-$
and $\frac32^+$. Hence, the potential candidates are
$\tilde\Lambda_{c2}(\frac52^-)$, $\Lambda_{c2}(\frac32^+)$,
$\tilde\Lambda'_{c1}(\frac32^+)$,
$\tilde\Lambda''_{c1}(\frac32^+)$ and
$\tilde\Lambda''_{c2}(\frac32^+)$. Ratios of $\Lambda_c(2940)$
partial widths are expected in HHChPT to be
 \be
 {\Gamma\left(\tilde\Lambda_{c2}(5/2^-)\to
[\Sigma_c^*\pi]_D\right)\over
\Gamma\left(\tilde\Lambda_{c2}(5/2^-)\to [\Sigma_c\pi]_D\right)}
&=&{7\over 2} \,{p_\pi^5(\Lambda_c(2940)\to\Sigma_c^*\pi)\over
p_\pi^5(\Lambda_c(2940)\to\Sigma_c\pi)}={7\over 2}\times
0.48=1.68\,,
\non \\
 {\Gamma\left(\Lambda_{c2}(3/2^+)\to
[\Sigma_c^*\pi]_P\right)\over \Gamma\left(\Lambda_{c2}(3/2^+)\to
[\Sigma_c\pi]_P\right)}&=&{1\over 5}
\,{p_\pi^3(\Lambda_c(2940)\to\Sigma_c^*\pi)\over
p_\pi^3(\Lambda_c(2940)\to\Sigma_c\pi)}={1\over 5}\times
0.65=0.13\,, \non \\
 {\Gamma\left(\tilde\Lambda'_{c1}(3/2^+)\to
[\Sigma_c^*\pi]_P\right)\over
\Gamma\left(\tilde\Lambda'_{c1}(3/2^+)\to
[\Sigma_c\pi]_P\right)}&=&
5\,{p_\pi^3(\Lambda_c(2940)\to\Sigma_c^*\pi)\over
p_\pi^3(\Lambda_c(2940)\to\Sigma_c\pi)}=5\times 0.65=3.25\,.
 \en
Since the predicted ratios differ significantly for different
$J^P$ quantum numbers, the measurements of the ratio of
$\Sigma_c^*\pi/\Sigma_c\pi$ will enable us to discriminate the
$J^P$ assignments for $\Lambda_c(2940)$.

For the charmed states $\Xi_c(2980)$ and $\Xi_c(3077)$, they could
be the first positive-parity excitations of $\Xi_c$ in viewing of
their large masses. Since the mass difference between the
antitriplets $\Lambda_c$ and $\Xi_c$ for
$J^P=\frac12^+,\frac12^-,\frac32^-$ is of order $180\sim 200$ MeV,
it is conceivable that $\Xi_c(2980)$ and $\Xi_c(3077)$ are the
counterparts of $\Lambda_c(2765)$ and $\Lambda_c(2880)$,
respectively, in the strange charmed baryon sector. As noted in
passing, the state $\Lambda_c(2765)^+$ could be an even-parity
excitation as the quark model \cite{Capstick} and the Skyrme model
\cite{Oh} suggest a $J^P=\frac12^+$ state with a mass 2742 and
2775 MeV, respectively. It is thus tempting to assign
$J^P=\frac12^+$ for $\Xi_c(2980)$ and $\frac52^+$ for
$\Xi_c(3077)$. Of course, the assignment of $J^P=\frac32^+$ is
also possible. The possible strong decays of the first
positive-parity excitations of the $\Xi_c$ states are summarized
in Table \ref{tab:Xic}. Since the two-body modes $\Xi_c\pi$,
$\Lambda_cK$, $\Xi'_c\pi$ and $\Sigma_cK$ are in $P$ ($F$) waves
and the three-body modes $\Xi_c\pi\pi$ and $\Lambda_cK\pi$ are in
$S$ ($D$) waves in the decays of $\frac12^+$ ($\frac52^+$), this
explains why $\Xi_c(2980)$ is broader than $\Xi_c(3077)$. Since
both $\Xi_c(2980)$ and  $\Xi_c(3077)$ are above the $D\Lambda$
threshold, it is important to search for them in the $D\Lambda$
spectrum as well.

\begin{table}[t]
\caption{Possible strong decays of the first positive-parity
excitations of the $\Xi_c$, where $L$ denotes the orbital angular
momentum of the light meson(s). The final state $\Sigma_c^*K$ is
kinematically allowed for $\Xi_c(3077)$.
 } \label{tab:Xic}
\begin{ruledtabular}
\begin{tabular}{c  l l l c}
$J^P$  & Diquark transition & $L$ & Decay channel & Final states \\
\hline
 ${1\over 2}^+$  & $0^+\to 1^++0^-$ & 1 &
 ${1\over 2}^+\to\{{1\over 2}^+,{3\over 2}^+\}+0^-$ & $\Xi'_c\pi,\Xi_c^*\pi,\Sigma_c^{(*)} K$
 \\
  & $0^+\to 0^++0^-+0^-$ & 0 & ${1\over 2}^+\to {1\over 2}^++0^-+0^-$ & $\Xi_c\pi\pi,\Lambda_cK\pi$
 \\
  & $1^+\to 0^++0^-$ & 1 & ${1\over 2}^+\to {1\over 2}^++0^-$ & $\Xi_c\pi,\Lambda_cK,D\Lambda$
 \\
  & $1^+\to 1^++0^-$ & 1 &
 ${1\over 2}^+\to\{{1\over 2}^+,{3\over 2}^+\}+0^-$ & $\Xi'_c\pi,\Xi_c^*\pi,\Sigma_c^{(*)} K$
 \\
  & $1^+\to 1^-+0^-$ & 0 &
 ${1\over 2}^+\to {1\over 2}^-+0^-$ & $\Xi_c(2790)\pi,\Lambda_c(2593)K$
 \\
  & $1^+\to 1^-+0^-$ & 2 &
 ${1\over 2}^+\to {3\over 2}^-+0^-$ & $\Xi_c(2815)\pi,\Lambda_c(2625)K$
 \\
  & $1^+\to 0^++0^-+0^-$ & 2 & ${1\over 2}^+\to {1\over 2}^++0^-+0^-$ & $\Xi_c\pi\pi,\Lambda_cK\pi$
 \\
  & $1^+\to 1^++0^-+0^-$ & 0 & ${1\over 2}^+\to {1\over 2}^++0^-+0^-$ & $\Xi'_c\pi\pi$
 \\
 & $1^+\to 1^++0^-+0^-$ & 2 & ${1\over 2}^+\to {3\over 2}^++0^-+0^-$ & $\Xi^*_c\pi\pi$
 \\
 \hline
 ${3\over 2}^+$  & $1^+\to 0^++0^-$ & 1 &
 ${3\over 2}^+\to {1\over 2}^++0^-$ & $\Xi_c\pi,\Lambda_cK,D\Lambda$
 \\
 & $1^+\to 1^++0^-$ & 1 &
 ${3\over 2}^+\to\{{1\over 2}^+,{3\over 2}^+\}+0^-$ & $\Xi'_c\pi,\Xi_c^*\pi,\Sigma_c^{(*)} K$
 \\
 & $1^+\to 1^-+0^-$ & 0 &
 ${3\over 2}^+\to {3\over 2}^-+0^-$ & $\Xi_c(2815)\pi,\Lambda_c(2625)K$
 \\
 & $1^+\to 1^-+0^-$ & 2 &
 ${3\over 2}^+\to\{{1\over 2}^-,{3\over 2}^-\}+0^-$ & $\Xi_c(2790,2815)\pi,\Lambda_c(2593,2625)K$
 \\
  & $1^+\to 0^++0^-+0^-$ & 2 & ${3\over 2}^+\to {1\over 2}^++0^-+0^-$ & $\Xi_c\pi\pi,\Lambda_cK\pi$
 \\
  & $1^+\to 1^++0^-+0^-$ & 0 & ${3\over 2}^+\to {3\over 2}^++0^-+0^-$ & $\Xi^*_c\pi\pi$
 \\
 & $1^+\to 1^++0^-+0^-$ & 2 & ${3\over 2}^+\to  \{{1\over 2}^+,{3\over 2}^+\}+0^-+0^-$ & $\Xi'_c\pi\pi,\Xi^*_c\pi\pi$
 \\
 & $2^+\to 1^++0^-$ & 1 &
 ${3\over 2}^+\to\{{1\over 2}^+,{3\over 2}^+\}+0^-$ & $\Xi'_c\pi,\Xi_c^*\pi,\Sigma_c^{(*)} K$
 \\
 & $2^+\to 1^-+0^-$ & 2 &
 ${3\over 2}^+\to\{{1\over 2}^-,{3\over 2}^-\}+0^-$ & $\Xi_c(2790,2815)\pi,\Lambda_c(2593,2625)K$
 \\
  & $2^+\to 0^++0^-+0^-$ & 2 & ${3\over 2}^+\to {1\over 2}^++0^-+0^-$ & $\Xi_c\pi\pi,\Lambda_cK\pi$
 \\
  & $2^+\to 1^++0^-+0^-$ & 2 & ${3\over 2}^+\to \{{1\over 2}^+,{3\over 2}^+\}+0^-+0^-$ & $\Xi'_c\pi\pi,\Xi^*_c\pi\pi$
 \\
 \hline
 $\frac52^+$ & $2^+\to 1^++0^-$ & 1 &
 ${5\over 2}^+\to {3\over 2}^++0^-$ & $\Xi_c^*\pi,\Sigma_c^{*} K$
 \\
 & $2^+\to 1^++0^-$ & 3 &
 ${5\over 2}^+\to\{{1\over 2}^+,{3\over 2}^+\}+0^-$ & $\Xi'_c\pi,\Xi_c^*\pi,\Sigma_c^{(*)} K$
 \\
 & $2^+\to 1^-+0^-$ & 2 &
 ${5\over 2}^+\to\{{1\over 2}^-,{3\over 2}^-\}+0^-$ & $\Xi_c(2790,2815)\pi,\Lambda_c(2593,2625)K$
 \\
  & $2^+\to 0^++0^-+0^-$ & 2 & ${5\over 2}^+\to {1\over 2}^++0^-+0^-$ & $\Xi_c\pi\pi,\Lambda_cK\pi$
 \\
  & $2^+\to 1^++0^-+0^-$ & 2 & ${5\over 2}^+\to \{{1\over 2}^+,{3\over 2}^+\}+0^-+0^-$ & $\Xi'_c\pi\pi,\Xi^*_c\pi\pi$
 \\
  & $3^+\to 0^++0^-$ & 3 &
 ${5\over 2}^+\to {1\over 2}^++0^-$ & $\Xi_c\pi,\Lambda_cK,D\Lambda$
 \\
 & $3^+\to 1^++0^-$ & 3 &
 ${5\over 2}^+\to\{{1\over 2}^+,{3\over 2}^+\}+0^-$ & $\Xi'_c\pi,\Xi_c^*\pi,\Sigma_c^{(*)}K$
 \\
 & $3^+\to 1^-+0^-$ & 2 &
 ${5\over 2}^+\to\{{1\over 2}^-,{3\over 2}^-\}+0^-$ & $\Xi_c(2790,2815)\pi,\Lambda_c(2593,2625)K$
 \\
  & $3^+\to 1^++0^-+0^-$ & 2 & ${5\over 2}^+\to \{{1\over 2}^+,{3\over 2}^+\}+0^-+0^-$ & $\Xi'_c\pi\pi,\Xi^*_c\pi\pi$
 \\
\end{tabular}
\end{ruledtabular}
\end{table}

\section{Conclusions}

Strong decays of charmed baryons are analyzed in the framework of
heavy hadron chiral perturbation theory  in which heavy quark
symmetry and chiral symmetry are synthesized. Our main conclusions
are the following:

\begin{itemize}
\item For $s$-wave charmed baryons, we use the channel
$\Sigma_c^{++}\to\Lambda_c^+\pi^+$ to fix the coupling constant
$g_2$. The value of $|g_2|=0.591\pm0.023$ are in good agreement
with the quark model expectation. The predictions for the strong
decays $\Sigma_c^*\to\Lambda\pi$ and $\Xi^*_c\to\Xi_c\pi$ are in
excellent agreement with experiment.

\item For $L=1$ orbitally excited baryons, two of the unknown
couplings, namely, $h_2$ and $h_{10}$, are determined from the
resonant $\Sigma_c^+\pi\pi$ mode produced in the $\Lambda_c(2593)$
decay and the width of $\Sigma_c(2800)$, respectively. The results
are $|h_2|=0.437^{+0.114}_{-0.102}$ and $|h_{10}|\lsim
(0.85^{+0.11}_{-0.08})\times 10^{-3}\,{\rm MeV}^{-1}$. Since the
two-pion system $\Lambda_c^+\pi\pi$ in $\Lambda_c(2593)^+$ decays
receives non-resonant contributions, our value for $h_2$ is
smaller than the previous estimates. Applying the quark model
relation $h_8^2=h_{10}^2$, predictions for the strong decays of
other $p$-wave charmed baryons such as
$\Lambda_c(2625)\to\Sigma_c\pi,\Lambda_c\pi\pi$,
$\Xi_c(2790)\to\Xi'_c\pi$ and $\Xi_c(2815)\to\Xi^*_c\pi$ are
presented in Table \ref{tab:strongdecayP}. Since the decays
$\Lambda_c(2593)\to\Lambda_c\pi\pi$ and
$\Lambda_c(2593)\to\Sigma_c\pi$ occur very close to the threshold,
they are very sensitive to the pion's mass and hence isospin
symmetry is violated, for example, $\Gamma(\Sigma_c^+\pi^0)\approx
2\Gamma(\Sigma_c^{++}\pi^-)$ and
$\Gamma(\Lambda_c^+\pi^0\pi^0)\approx
\Gamma(\Lambda_c^+\pi^+\pi^-)$ in $\Lambda_c(2593)$ decays.

\item We have examined the first positive-parity excited charmed
baryons. We conjecture that the state $\Lambda_c(2880)$ with
$J^P=\frac52^+$ is an admixture of $\Lambda_{c2}(\frac52^+)$ with
and $\tilde\Lambda''_{c3}(\frac52^+)$; both are $L=2$ orbitally
excited states. Potential models suggest the possible allowed
$J^P$ numbers of the $\Lambda_c(2940)^+$ to be $\frac52^-$ and
$\frac32^+$. We have demonstrated that the measurements of the
ratio of $\Sigma_c^*\pi/\Sigma_c\pi$ will enable us to
discriminate the $J^P$ assignments for $\Lambda_c(2940)$. We
advocate that the $J^P$ quantum numbers of $\Xi_c(2980)$ and
$\Xi_c(3077)$ are $\frac12^+$ and $\frac52^+$, respectively. Under
this $J^P$ assignment, it is easy to understand why $\Xi_c(2980)$
is broader than $\Xi_c(3077)$.

\end{itemize}

\vskip 2.0cm \acknowledgments  We are grateful to  Roman Mizuk for
valuable discussions. This research was supported in part by the
National Science Council of R.O.C. under Grant Nos.
NSC95-2112-M-001-013 and  NSC95-2112-M-033-013.

\newpage
\newcommand{\bi}{\bibitem}

\end{document}